\documentclass[12pt,a4]{article}

\usepackage{amsthm,amsmath,latexsym,amssymb,amsfonts,amscd}
\usepackage{graphics,lscape,fancyhdr,array,stmaryrd,euscript}
\usepackage{ulem}
\usepackage{ifthen,empheq}
\usepackage{sidecap}
\usepackage{ifpdf}
\usepackage{verbatim}
\usepackage{color}
\usepackage{tikz-cd}
\usetikzlibrary{cd}
\usepackage{relsize,slashed}
\numberwithin{equation}{section}
\usepackage{hyperref}

\newcommand{\bep}{\begin{picture}}
\newcommand{\eep}{\end{picture}}

\newcounter{YoungHeight}\newcounter{YoungWidth}

\newcounter{Mul1}\newcounter{Mul2}\newcounter{Mul3}\newcounter{Mul4}
\newcounter{A0}\newcounter{A1}\newcounter{A2}
\newcounter{B3}
\newcounter{C3}\newcounter{C4}
\newcounter{D1}\newcounter{D2}\newcounter{D3}
\newcounter{T0}\newcounter{T1}

\newlength{\txtHShift}

\newlength{\txtWidth}

\newcommand{\HalfLength}[2]{\setcounter{Mul1}{#1}\setcounter{Mul2}{#1}\addtocounter{Mul1}{\value{Mul2}}\addtocounter{Mul1}{\value{Mul2}}%
\addtocounter{Mul1}{\value{Mul2}}\addtocounter{Mul1}{\value{Mul2}}\setcounter{#2}{\value{Mul1}}}

\newcommand{\Add}[3]{\setcounter{#1}{#2}\addtocounter{#1}{#3}}

\newcommand{\Length}[1]{#10}
\newcommand{\YoungScale}{}%\unitlength=0.35mm}

\newcommand{\shiftedText}[2]{{\hspace{#1}#2}}
\newcommand{\calcHShift}[1]{\settowidth{\txtWidth}{#1}\setlength{\txtHShift}{-0.5\txtWidth}}

\newcommand{\TextCenter}[3]{{\HalfLength{#2}{T0}%
\HalfLength{#3}{T1}\addtocounter{T1}{-3}\calcHShift{#1}%
\put(\value{T0},\value{T1}){\shiftedText{\txtHShift}{#1}}}}

\newcommand{\TextCenterB}[3]{{\calcHShift{#1}\HalfLength{#2}{T0}\Add{T1}{\Length{#3}}{-7}\put(\value{T0},\value{T1}){\shiftedText{\txtHShift}{#1}}}}

\newcommand{\BlockA}[2]{{\YoungScale\bep(\Length{#1},\Length{#2}){\Add{A1}{#1}{1}\Add{A2}{#2}{1}}%
\multiput(0,0)(10,0){\value{A1}}{\line(0,1){\Length{#2}}}\multiput(0,0)(0,10){\value{A2}}{\line(1,0){\Length{#1}}}%
\setcounter{YoungHeight}{\Length{#2}}\setcounter{YoungWidth}{\Length{#1}}\eep}}

\newcommand{\RectT}[3]{\bep(\Length{#1},\Length{#2})\put(0,0){\line(1,0){\Length{#1}}}\put(0,0){\line(0,1){\Length{#2}}}%
\put(\Length{#1},\Length{#2}){\line(-1,0){\Length{#1}}}\put(\Length{#1},\Length{#2}){\line(0,-1){\Length{#2}}}#3{#1}{#2}\eep}

\newcommand{\RectARowUp}[2]{{\bep(\Length{#1},10)\put(0,0){\RectT{#1}{1}{\TextCenterB{#2}}}\eep}}

\newcommand{\YoungA}{\BlockA{1}{1}}

\usepackage{setspace}

 \usepackage[numbers,sort&compress]{natbib}
 \usepackage[nottoc]{tocbibind}
\newcommand{\pl}{\partial}
\newcommand{\besubeqs}{\begin{subequations}}
\newcommand{\esubeqs}{\end{subequations}}

\newcommand{\maxim}[1]{\textsc{#1}}

\renewcommand{\tilde}{\widetilde}

 \newcommand{\bref}[1]{\textbf{\ref{#1}}}
\newcommand{\p}[1]{|#1|}
\newcommand{\gh}[1]{\mathrm{gh}(#1)}

\newcommand{\dd}{\partial}
\renewcommand{\d}{\partial}

\newcommand{\tensor}{\otimes}

\renewcommand{\geq}{\,{\geqslant}\,}

\newcommand{\binner}[2]{%
  {\langle}\kern-4.15pt{\langle}#1{,}\,#2{\rangle}\kern-4.15pt{\rangle}}
\newcommand{\commut}[2]{[#1{,}\,#2]}

\newcommand{\qcommut}[2]{[#1{,}\,#2]_\star}

\newcommand{\half}{\mathchoice{%
    \ffrac{1}{2}}{\frac{1}{2}}{\frac{1}{2}}{\frac{1}{2}}}

\newcommand{\ffrac}[2]{\raisebox{.5pt}%
  {\footnotesize$\displaystyle\frac{#1}{#2}$}\kern1pt}

 \newcommand{\red}{\mathrm{red}}

\newcommand{\dl}[1]{\mathchoice{\ffrac{\dd}{\dd #1}}{\frac{\dd}{\dd
      #1}}{\ffrac{\dd}{\dd #1}}{\ffrac{\dd}{\dd #1}}}
 \newcommand{\fC}{\mathbb{C}}
 \newcommand{\fR}{\mathbb{R}}
 \newcommand{\fZ}{\mathbb{Z}}
 \def\cH{\mathcal{H}}
 \def\cS{\mathcal{S}}
 \def\BGST{Barnich:2004cr}

\usepackage{etoolbox}
\preto\subequations{\ifhmode\unskip\fi}

\newcommand{\Sm}{{\ensuremath{\mathbf{m}}}}

\newcommand{\fud}[2]{{}^{#1}{}_{#2}\,}
\newcommand{\fdu}[2]{{}_{#1}{}^{#2}\,}

\newcommand{\mph}{{\theta}}
\newcommand{\Mph}{{\theta}}

\newcommand{\tildef}{{h}}
\newcommand{\hrepl}{{\mathrm{h}}}
\newcommand{\frepl}{{\mathrm{f}}}
\newcommand{\hsb}{{\mathrm{hs}_B}}
\newcommand{\Cone}{{\mathsf{C}}}
\newcommand{\kkk}{\mathfrak{k}}
\newcommand{\uA}{{{\underline{A}}}}
\newcommand{\uB}{{{\underline{B}}}}
\newcommand{\hsel}{{\mathsf{a}}}
\newcommand{\hselb}{{\mathsf{b}}}
\newcommand{\hselc}{{\mathsf{c}}}

\usepackage[toc]{multitoc}
\usepackage{tocloft}

\begin{document}
%%%%%%%%%%%%%%%%%%%%%%%%%%%%%%%%%%%%%%%%%%%%%%%%%%%%%%%%%%%%%
\pagenumbering{gobble}
\hfill
\vspace{-1.5cm}
\begin{flushright}
    {LMU-ASC 17/18}
\end{flushright}
\vskip 0.01\textheight
\begin{center}
{\Large\bfseries 
Type-B Formal Higher Spin Gravity}

%\vspace{0.4cm}

\vskip 0.03\textheight

Maxim Grigoriev${}^{a,b}$ and Evgeny Skvortsov${}^{c,b}$

\vskip 0.03\textheight

{\em ${}^{a}$ Arnold Sommerfeld Center for Theoretical Physics\\
Ludwig-Maximilians University Munich\\
Theresienstr. 37, D-80333 Munich, Germany}\\

\vspace{5pt}
{\em$^b$ Lebedev Institute of Physics, \\
Leninsky ave. 53, 119991 Moscow, Russia}\\

\vspace{5pt}
{\em$^c$  Albert Einstein Institute, \\
Am M\"{u}hlenberg 1, D-14476, Potsdam-Golm, Germany}

\end{center}

\vskip 0.02\textheight

\begin{abstract}
We propose non-linear equations for the formal Type-B Higher Spin Gravity that is dual to the free fermion or to the Gross-Neveu model, depending on the boundary conditions. The equations are directly obtained from the first principles: the gauge invariance of the CFT partition function on an arbitrary background for single-trace operators. We also get equations describing propagation of certain mixed-symmetry fields over higher spin flat backgrounds.
\end{abstract}
{ \footnotesize
\tableofcontents }

\newpage
%%%%%%%%%%%%%%%%%%%%%%%%%%%%%%%%%%%%%%%%%%%%%%%%%%%%%%%%%%%%%
\section{Introduction}
%%%%%%%%%%%%%%%%%%%%%%%%%%%%%%%%%%%%%%%%%%%%%%%%%%%%%%%%%%%%%
\pagenumbering{arabic}
\setcounter{page}{2}
Higher Spin Gravities are theories featuring massless fields with spin greater than two on top of the spin-two graviton. When considered on AdS background the masslessness of the bulk fields implies \cite{Sundborg:2000wp,Sezgin:2002rt,Klebanov:2002ja} that the AdS/CFT dual operators are conserved tensors, which in $d>2$ is a clear signature of free CFT's \cite{Maldacena:2011jn,Alba:2013yda,Boulanger:2013zza,Alba:2015upa}. On general grounds it is expected that any free CFT should have a higher spin dual. The dual is then a theory that computes correlation functions of single-trace operators on the CFT side as Witten diagrams on the AdS side. The problem we address in the paper is how to reconstruct the dual higher spin gravity given some free CFT. Solving this problem also gives an access to more interesting dualities that can be obtained by imposing different boundary conditions, see e.g. \cite{Klebanov:1999tb,Klebanov:2002ja,Sezgin:2003pt,Leigh:2003gk,Giombi:2011ya,Bekaert:2012ux}.

One special feature of free CFT's is that the global conformal symmetry gets extended to the global infinite-dimensional higher spin symmetry. The operators of interest for higher spin AdS/CFT are various bilinears, single-trace operators, most of which are conserved tensors responsible for the higher spin symmetry. It turns out that the higher spin symmetry is powerful enough as to fix unambiguously the correlators of the single-trace operators  \cite{Maldacena:2011jn,Alba:2013yda,Boulanger:2013zza,Alba:2015upa}. The single-trace operators are dual to the fields of a higher spin gravity. Bearing in mind the conceptual simplicity of the higher spin AdS/CFT duality, it would be important to directly map any given free CFT data to a higher spin gravity. 

The key observation is that the unambiguity of the generating functional of correlators can also be understood by turning on sources for the single-trace operators and studying gauge invariance of the partition function as a functional of these sources as those that couple to conserved tensors are gauge fields. The gauge symmetries of the sources encode Ward identities of the higher spin symmetry. Therefore, the unambiguity of the correlators can be understood from a purely classical theory of background fields:
the infinite-dimensional non-abelian gauge symmetry of sources completely fixes the effective action. It is worth stressing that this phenomenon takes place only for CFT's with higher spin symmetry. In the usual low spin case, for instance, the generating functional $W[g]$ of stress-tensor correlation functions should be invariant under diffeomorphisms and (up to an anomaly) under Weyl rescalings, which does not allow one to reconstruct $W[g]$ in $d>2$. On the contrary, knowing the non-abelian symmetries of the higher spin sources $h_s$, $s=0,1,2,...$, allows one to reconstruct $W[h_s]$, which is then can be seen to be a generating functional of single-trace operators of a free CFT. 

Our strategy is to explore the theory of higher spin background fields. We first show how to put a CFT on an arbitrary higher spin background while preserving the invariance of the partition function (effective action for background fields) and then uplift the theory of background fields to $AdS_{d+1}$ where these background fields turn into boundary values. As different from other approaches, no perturbative expansion of any kind is needed, the theory is reconstructed in one shot. This strategy has already been applied in \cite{Bekaert:2017bpy} to the simplest higher spin gravity, Type-A, which was reconstructed from the free scalar CFT.\footnote{It is also worth mentioning somewhat implicit proposal~\cite{Bekaert:2013zya} of the AdS dual of higher order nonunitary scalar CFT, $\square^k\phi=0$, as well as the general framework developed in~\cite{Grigoriev:2006tt,Grigoriev:2011gp,Bekaert:2012vt}. } Historically, the $4d$ Type-A theory was originally proposed by Vasiliev \cite{Vasiliev:1988sa,Vasiliev:1990cm,Vasiliev:1999ba} as intrinsically AdS theory before AdS/CFT era began \cite{Maldacena:1997re,Witten:1998qj,Gubser:1998bc}.

In the present paper we strengthen the universality of the method, its relation to background fields\footnote{By the theory of background fields we mean non-infinitesimal sources together with gauge symmetries thereof that make the effective action gauge invariant.} on the CFT side and apply it to reconstruct the Type-B higher spin gravity that is a dual of the free fermion CFT. Upon changing the boundary conditions, the same theory should also be dual to the Gross-Neveu model, where higher spin symmetry is broken by $1/N$ effects, \cite{Sezgin:2003pt,Leigh:2003gk}.

In slightly more detail, the general higher spin AdS/CFT story for any free vectorial CFT is as follows. Let us discuss the case of the free scalar CFT. Given a free conformal field $\phi$, $\square \phi=0$, one can detect an infinite-dimensional extension of the conformal symmetry, the higher spin symmetry, by constructing higher rank conserved tensors:
\begin{align}
    J_s\equiv J_{a_1...a_s}&= \bar\phi \pl_{a_1}...\pl_{a_s}\phi +\text{more terms to ensure conservation}\,,
\end{align}
where the stress-tensor is $J_{s=2}$. The conserved tensors $J_s$, via Noether theorem, give rise to higher spin algebra transformations $\delta \phi$ that are various differential operators with polynomial coefficients that map solutions $\square \phi=0$ to solutions, the conformal transformations being represented by operators with no more than one derivative \cite{Nikitin1991,Eastwood:2002su}. Together with $J_0=\bar\phi \phi$, higher spin currents $J_s$ make a full list of single-trace operators. Therefore, the AdS dual description should feature one scalar field dual to $J_0$ and gauge fields dual to $J_{s>0}$. Sources for $J_s$ are conformal higher spin fields $h^s\equiv h^{a_1...a_s}(x)$ subject to the gauge transformations
\begin{align}\label{CHSgauge}
    \delta h^{a_1...a_s}&= \pl^{a_1} \xi^{a_2...a_s}+\text{permutations}+\text{higher orders}\,,
\end{align} 
reflecting the conservation of $J_{s>0}$. The nontrivial problem is to extend $h^{a_1...a_s}$ beyond the infinitesimal level and promote them to background fields. This requires counterterms of the form $h^2 \bar\phi\phi$ and higher as well as corrections to the gauge transformations \eqref{CHSgauge} that are of order $h\xi$ and higher. A priori it is not obvious how to tackle this problem for higher spin background fields. 

For example, in the lower spin case, the source for the global symmetry current, say $u(M)$, would be vector potential $\delta A_c=\pl_c \xi$ and, as is well-known, one has to add the counterterm $A^2\bar\phi\phi$ coming from $D_A\bar\phi D_A\phi$ in order to make the effective action $W[A]$ gauge invariant, while the gauge transformations should be extended to $\delta A_c=\pl_c \xi+[A_c,\xi]$. Similarly, coupling to conformal metric $T_{ab }g^{ab}$ entails an infinite number of counterterms that can be resummed to the usual kinetic term $\sqrt{g} \pl\bar\phi \pl\phi$ and $\sqrt{g} R\bar\phi\phi$. These two examples correspond to the $s=1,2$ sources $h^{s}$ and are particular cases of the problem on how to promote all of the $h^{s}$ to background fields. The final result contains counterterms of all orders $(h)^k\,\bar\phi\phi$ in the background fields and corrections to the gauge transformations \eqref{CHSgauge} of all orders as well. 

The first part of the problem on how to couple to a higher spin background was solved in \cite{Segal:2002gd} (see also~\cite{Tseytlin:2002gz,Bekaert:2010ky,Grigoriev:2016bzl}). It turns out that all the counterterms can be resummed and the CFT partition function with sources for the single-trace operators turned on is simply
\begin{align}\label{scalexam}
   e^{W[H]}&=\int D\phi\,D\bar\phi\, e^{ -\int \bar\phi H \phi}\,, &&W[H]=- \mathrm{tr} \log H \,,
\end{align}
where an advantage has been taken of the fact that the terms $h^{a_1...a_s}(x)J_{a_1....a_s}$ can be represented as some operators sandwiched in between $\bar\phi$, $\phi$ and then they can be combined with the kinetic term $-\pl^2$ as to form certain operator $H$. For infinitesimal sources $W[H]$ can be expanded over $-\pl^2$ as to give the expected generating functional of currents $J_s$. When written as \eqref{scalexam} the functional is clearly invariant for non-infinitesimal sources, i.e. when $h^{a_1\ldots a_s}$ represent a non-trivial higher spin background. The transformations for $h^{a_1\ldots a_s}$ can be read off from naive\footnote{In even dimensions $W[h^s]$ has a local part, the (higher spin) conformal anomaly, \cite{Segal:2002gd,Tseytlin:2002gz}.} gauge symmetries of the action $\int \bar \phi H \phi$: $\delta \phi=-U\phi$ and $\delta H=HU+U^\dag H$, where $U$ is any operator in $x$ and $\pl$. 

The seeming simplicity of the solution is deceptive, the devil being in how to relate the components of $H$ to the background fields $h^{a_1\ldots a_s}$. For example, it takes a non-linear redefinition to recover $\sqrt{g}R\bar\phi\phi$. Nevertheless, that the theory of background fields, i.e. $H$ with its gauge transformations, is known is sufficient for our purposes.  

Dual higher spin gravity is supposed to provide another way to compute the effective action {$W[h^s]$}. The fundamental fields of the higher spin theory are bulk fields whose boundary values correspond to $h^{s}$. As it has already been stressed, the special feature of higher spin symmetry is that the infinite-dimensional non-abelian gauge symmetries of background fields $h^{s}$ unambiguously fix $W[h^s]$. The global higher spin symmetry of a free CFT should become gauge symmetry of its bulk dual. The gauge symmetries of $h^{s}$ are closely related to those of the dual theory: at the very least the near boundary analysis of the dual theory should reveal the same $h^{s}$. In fact, our basic conclusion is that they are the same: the dual theory at the classical level can be extracted directly from the gauge symmetries of $h^{s}$.

There are three technical steps that allow us to make the general ideas above more precise. Firstly, we uplift a given free CFT to the ambient space $\mathbb{R}^{d,2}$ where the conformal space is embedded as projective cone, $X^2=0$, $X\sim \lambda X$, $\lambda>0$. The advantage is that the conformal symmetry is now manifest. Secondly, the theory of background higher spin fields can also be uplifted to the ambient space.  
For the case of a free scalar field, a theory of background fields turns out to be equivalent to three first class constraints on the ambient space that form $sp(2)$. This is the same $sp(2)$ that underlies the Fefferman-Graham construction \cite{FG,GJMS}. The third and the main step is that the same constrained system can be considered on the hyperboloid $X^2=-1$, where it can be shown to describe an appropriate multiplet of higher spin fields.

These steps are performed using the combination of various techniques: a version~\cite{Barnich:2006pc,Bekaert:2009fg,Grigoriev:2011gp} of the conventional ambient space formalism, parent formulation method which allows to implement ambient construction in the fiber rather than in space-time~\cite{Barnich:2004cr,Grigoriev:2006tt}, and the approach to boundary values employing both the ambient and the parent techniques~\cite{Bekaert:2012vt,Bekaert:2013zya}.

Our main goal in the present paper is to show that the same reasoning works fine for the underlying free CFT being a Dirac fermion. The higher spin dual of the free fermion vector model in any $d$ has been already dubbed Type-B in the literature \cite{Sezgin:2003pt}, even though the model has not yet been constructed in any detail in $d>3$. The single-trace operators and thereby the background fields reveal a trickier pattern: tensors with mixed-symmetry show up. While totally-symmetric higher spin fields relevant for the Type-A duality have been thoroughly studied, mixed-symmetry fields resisted so far any attempt to introduce interactions. In particular, there have been no results on Type-B theory.\footnote{See, however, \cite{Beccaria:2014xda,Giombi:2014yra,Giombi:2016pvg,Gunaydin:2016amv} for some one-loop determinants and \cite{Alkalaev:2010af,Boulanger:2011qt,Boulanger:2011se,Boulanger:2012dx} for some cubic interactions.} The main statements are: (i) theory of background fields for the singe-trace operators in the free fermion model can be described as $osp(1|2)$ constraints on ambient phase space;\footnote{$osp(1|2)$ constraints in ambient space were also considered in \cite{Bonezzi:2015bza} in somewhat similar context.} (ii) on the hyperboloid $X^2=-1$ this system gives formally consistent and gauge invariant equations; (iii) when linearized over $AdS_{d+1}$ the equations describe the right spectrum of massless and massive fields; (iv) over the maximally symmetric, higher spin flat, backgrounds we show that the first order corrections yield the expected Hochschild cocycle of the higher spin algebra, which is the only non-trivial data that goes beyond the higher spin algebra itself. This also gives equations that describe how fields of Type-B theory propagate over any flat background.

The outline of the paper is as follows. In section \bref{sec:fermion-mink} we describe the theory of background fields for single-trace operators of the free fermion CFT. It is uplifted to the ambient space in section \bref{sec:ambient}. In section \bref{sec:ads-free} we discuss free fields of the Type-B theory in $AdS_{d+1}$ that are dual to the background fields. The proposal for the Type-B theory is in section \bref{sec:typeb}, where its various properties and limits are discussed. The relation to the formal deformations and Hochschild cohomology is discussed in \bref{sec:fda}. The conclusions and discussion are in section~\bref{sec:conclusions}.

%%%%%%%%%%%%%%%%%%%%%%%%%%%%%%%%%%%%%%%%%%%%%%%%%%%%%%%%%%%%%
\section{Fermion on Higher Spin Background}
\label{sec:fermion-mink}
%%%%%%%%%%%%%%%%%%%%%%%%%%%%%%%%%%%%%%%%%%%%%%%%%%%%%%%%%%%%%
As it was briefly explained in the introduction, our strategy is to put a fermion on a nontrivial background in which we turn on sources for all 'single-trace' operators. By single-trace operators we mean all quasi-primary operators that are bilinear in the fermion. Majority of these operators are conserved tensors and for that reason the sources are gauge fields. The latter makes nontrivial a problem of extending infinitesimal sources to background fields. In particular, the linearized gauge symmetry responsible for the conservation of tensors in flat space (trivial background) has to be deformed into some sort of covariant conservation on a higher spin background. Among the background fields there is a usual $u(1)$ gauge field, conformal metric and infinitely many sources for higher spin conserved tensors. The sought for deformed symmetries will have to be infinite-dimensional and non-abelian, mixing sources of different spins. Surprisingly, this problem is very easy to solve, as was proposed in~\cite{Segal:2002gd} in the case of scalar field, see also~\cite{Bekaert:2010ky,Grigoriev:2016bzl}.

First of all, let us remind that the set of single-trace operators consists of irreducible tensors with the symmetry of hook-shaped Young diagrams, with the indices in the column supplied by gamma-matrices and indices in the row by the space-time derivatives:
\begin{align}\label{singletrace}
    J_{s,q}\equiv J_{a_1...a_s,m_1...m_q}&=\bar{\psi} \gamma_{a_1 m_1...m_q} \pl_{a_2}...\pl_{a_{s}} \psi+... && J_{s,q}\sim\parbox{60pt}{{\bep(60,50)\unitlength=0.38mm%
    \put(0,40){\RectARowUp{6}{$s$}}%
    \put(0,0){\RectT{1}{4}{\TextCenter{$q$}}}\eep}}
\end{align}
Here $\gamma^{c_1...c_q}$ are anti-symmetrized products of $\gamma$-matrices, $\gamma^a\gamma^b + \gamma^b\gamma^a=2\eta^{ab}$. 
Explicit expressions for the currents are rather cumbersome due to the need to maintain Young symmetry, tracelessness and conservation, see e.g. \cite{Anselmi:1999bb,Alkalaev:2012rg,Giombi:2017rhm} for some examples. 

Among the currents one finds the totally-symmetric conserved tensors, as the $q=0$ case. In addition to conserved tensors there is a number of 'massive' $q$-forms
\begin{align}\label{masssivepforms}
    J_{m_1...m_q}&= \bar{\psi} \gamma_{m_1}...\gamma_{m_q} \psi\,, && q=0,2,3,4,...\,,
\end{align}
that are dual to massive anti-symmetric fields in $AdS_{d+1}$. Note that all single-trace operators can, in principle, be extracted from a set of bilocal generating functions, see e.g. \cite{Alkalaev:2012rg}:
\begin{align}\label{genfunc}
    \bar\psi(x-y) \gamma^{c_1...c_q} \psi(x+y)\,.
\end{align}
This set is over-complete in the sense it contains descendants and redundant operators. 

Let us consider a free Dirac fermion $\psi$ and view the equations of motion
\begin{align}
S_0&= \int \bar{\psi} (i \gamma\cdot \pl_x) \psi && \Longrightarrow && i\gamma^a \d_a \psi=0    
\end{align}
as a physical state condition in the underlying quantum constrained system. The constraints are first class and read
\begin{equation}\label{freeconstraints}
    -\d_x\cdot\d_x\,, \quad i\gamma\cdot \d_x\,,
    \qquad \qquad \commut{i\gamma\cdot\d_x}{i\gamma\cdot\d_x}=-2\d_x\cdot \d_x\,,
\end{equation}
where the brackets denote super-commutator, with the Grassmann degree being the standard $\fZ_2$-grading in the Clifford algebra of $\gamma$-matrices, so that in the above relations it is the anti-commutator. The constraints are operators acting in the space of "wave functions" with values in the Dirac spinor representation. Here and below we exploit the analogy between free fields and first-quantized particle models.

Now suppose we would like to add at least infinitesimal sources for all operators \eqref{singletrace}, i.e. couplings of the type
\begin{align}\label{couplingirr}
    S_1=\sum_{s,q}\int J_{a_1...a_s,c_1...c_q} \left(\tildef^{a_1...a_s,c_1...c_q}+\pl^{a_1}\xi^{a_2...a_s,c_1...c_q}+\text{perm}-\text{traces}\right)\,,
\end{align}
where we also indicated the desired gauge invariance. Here, the sources $\tildef^{s,q}$ have the same algebraic properties as $J_{s,q}$, i.e. they are traceless and have the Young symmetry depicted in \eqref{singletrace}. Since most of $J_{s,q}$ are conserved tensors (save for \eqref{masssivepforms}), the sources $\tildef^{s,q}$ enjoy a gauge symmetry. 

At this stage the sources in~\eqref{couplingirr} are infinitesimal. If they are not, the currents $J_{s,q}$ are not conserved: the invariance of the effective action $W[\tildef]$ can be restored upon adding appropriate counterterms $\bar\psi(\tildef...\tildef)\psi$ to $S_1$ and correcting the gauge symmetry $\delta\tildef$ by non-linear terms. This also results in nonlinear deformation of the conservation condition for the currents. As we explained in the Introduction, an elegant way to introduce sources so that the total action $S_0+S_1$ remains gauge invariant even for non-infinitesimal sources was proposed in~\cite{Segal:2002gd} in the case of scalar field (see also~\cite{Grigoriev:2016bzl} for a recent discussion of curved backgrounds and nonlinear gauge symmetries). Now we are going to apply the same strategy to the Dirac fermion.

The crucial step is to realize that all derivatives of $\bar{\psi}$ that are hidden in $J_{s,q}$ can be integrated by parts and the coupling can be written in the following suggestive form
\begin{align}\label{nicecoupling}
   S_1=i\int \bar{\psi}\, \frepl(\gamma, \pl_x|x) \psi\,,
\end{align}
where $\frepl$ is a function of $\gamma$ and $\pl$, whose Taylor coefficient can depend on $x$: 
\begin{align}
    \frepl(\gamma, \pl_x|x)&= \sum_{s,q} \frepl^{a_1...a_s|c_1...c_q}(x) \gamma_{c_1...c_q}\pl_{a_1}....\pl_{a_s}\,.
\end{align}
Here $\frepl^{s,q}$ are not irreducible tensors anymore. These coefficients are related to the source $\tildef^{s,q}$, but it makes sense not to work at this stage with the operators that are quasi-primary and to extend the base of local bilinear operators to the one covered by generating functions \eqref{genfunc}. It is clear that $S_1$ encodes the usual lower spin couplings like Yukawa, electro-magnetic and gravitational ones as well as couplings to higher spin sources. In what follows we refer to $\frepl$ as to background fields to stress that they are not required to be infinitesimal.

The full action of the conformal Dirac field coupled to all sources can be written as\footnote{Note that the usual minimal coupling of fermion to gravity, $|e| \bar{\psi} \gamma^a e_a^\Sm (\pl_\Sm+\omega^{a,b}_\Sm \tfrac18[\gamma_a,\gamma_b])\psi$, is certainly a part of \eqref{nicecoupling} upon appropriate identification of fields, see also the comment below about general covariance. The Taylor coefficients of $\frepl$ should be understood as $\frepl\fdu{a_1...a_s}{\Sm_1...\Sm_q}$, i.e. as tensors of the fiber Lorentz group $O(d-1,1)$ and transforming as coefficients of differential operators in the $\Sm$'s. In the present paper we do not dwell on the global geometry issues.}
\begin{equation}
\label{spinor-lag}
S=S_0+S_1=\int d^d x \,{\bar\psi}{(i\gamma\cdot \d_x+ i\frepl)\psi}\,,
\end{equation}
where it is useful to absorb the kinetic term into $F=i\gamma\cdot \d_x+ i\frepl$.

To describe sources in terms of the constrained system \eqref{freeconstraints}, we should also allow for a deformation $\hrepl=\hrepl(\gamma,\pl_x|x)$ of the $-\pl^2_x$ constraint, which can altogether be packed into $H=-\pl^2_x+\hrepl$. In these terms the equations of motion for the Dirac field are identified as physical state conditions: 
\begin{align}\label{FHsystem}
   F\psi&=(i\gamma\cdot \d_x+ i\frepl)\psi=0\,,&
    H\psi&=(-\pl^2_x+\hrepl)\psi=0\,.
\end{align}
If we insist that the constraint algebra is unchanged we find
\begin{align}\label{constraints}
[F,F]&= 2H && \Longleftrightarrow &&  2\commut{\gamma\cdot\d_x}{ \frepl}
+\commut{ \frepl}{ \frepl}= -2\hrepl\,,
\end{align}
so that, as expected, $\hrepl$ is determined by $\frepl$ and $H$ does not introduce any new independent sources.

There is a natural gauge symmetry which acts on both $\psi$ and $F$ and leaves the Lagrangian~\eqref{spinor-lag} unchanged. Indeed, consider
\begin{align}\label{Fpsigauge}
    \delta F&=[\epsilon,F]+\{\alpha,F\} +\{\beta, H\}\,, &
    \delta \psi&= \epsilon \psi -\alpha \psi -F\beta \psi\,,
\end{align}
where the three gauge parameters $\epsilon$, $\alpha$ and $\beta$ are functions of the same type as $F$, i.e. they depend on $\gamma$, $\pl_x$ and $x$. The reality of the action implies $\epsilon^\dag=-\epsilon$, $\alpha^\dag=\alpha$ and $\beta^\dag=\beta$. The $\epsilon$-symmetry is a usual gauge transformation, while $\alpha$ and $\beta$ are responsible for reparameterization of the constraints, which we discuss in the next section.

To study gauge symmetry for $F$, $H$ it is convenient to work in terms of symbols rather than operators. To this end we introduce variables $\mph^a$, $p_a$ associated with $\gamma^a$ and $\d_a$, respectively. The algebra of operators is then the tensor product of Weyl algebra and Clifford algebra. Seen as a $\star$-product algebra of symbols it can be identified as an algebra freely generated by $x^a$, $p_b$, $\mph^a$ modulo the following relations:
\begin{align}
\qcommut{x^a}{p_b}&=\delta^a_b\,, &
\qcommut{\mph^a}{\mph^b}&=2\eta^{ab}\,,
\end{align}
where $\qcommut{A}{B}\coloneqq A\star B-(-1)^{\p{A}\p{B}} B\star A$ denotes $\star$-product super-commutator, with the Grassmann degree assigned according to $\p{x^a}=\p{p_a}=0$ and $\p{\theta^a}=1$. The linearized gauge symmetries of $\frepl$ read as
\begin{equation}\label{linsym}
    \delta \frepl= -(p\cdot \d_\mph-\mph\cdot \d_x)\epsilon+2\alpha (p\cdot\mph)+2\beta (p\cdot p)+... \,,
\end{equation}
where we dropped the terms with derivatives of $\alpha$ and $\beta$. 

Now we would like to sketch the argument for why $\frepl$ is equivalent to the irreducible set of sources we introduced in \eqref{couplingirr}. The sources in \eqref{couplingirr} are irreducible: they have definite Young symmetry and are traceless. The symmetry with $\alpha$ and $\beta$ accompanied by appropriate field redefinitions allows one to set all traces to zero. Indeed, any trace is proportional to $(p\cdot\mph)$ or $(p\cdot p)$. As we demonstrate in Appendix \bref{sec:symmbase} in a more general setting the $\epsilon$-symmetry can be used to reach the gauge condition $(\mph\cdot\d_p) \frepl=0$, which is equivalent to the required Young symmetry for components. Then, for $\tildef=(p\cdot\d_\mph) \frepl$ we find 
\begin{equation}
    \delta \tildef= \Pi((p\cdot\d_x) \epsilon)\,,
\end{equation}
where $\epsilon$ is assumed to satisfy $(p\cdot\d_\mph) \epsilon=0$ and $\Pi$ denotes projector onto the traceless part. Therefore, we see that, at the linearized level, i.e. for infinitesimal sources, the extended gauge symmetries that act on $\frepl$ allow us to get rid of the redundant components and reproduce the coupling \eqref{couplingirr} to the irreducible sources $\tildef^{s,q}$.

Therefore, fluctuations $\frepl$ of $F$ over $F_0=i(\gamma\cdot p)$ are equivalent to the set of the off-shell conformal hook-type sources \eqref{couplingirr} with the required gauge transformations. This also solves the original problem: how to put a fermion on non-infinitesimal higher spin background. Indeed, the gauge symmetries \eqref{Fpsigauge} feature a non-abelian field-dependent part hidden in $[\epsilon, F]$ that is needed  to make the action invariant. Note, however, that even at the linearized level the proof that \eqref{linsym} is equivalent to \eqref{couplingirr} involves a chain of gauge fixings and redefinitions. These redefinitions become more complicated if we want to extract the non-linear corrections to the gauge symmetries \eqref{couplingirr} out of \eqref{Fpsigauge}.   

An important remark is that the action \eqref{spinor-lag} does not seem to respect general covariance. In fact this is not so. Diffeomorphisms are a part of the gauge symmetry: $\epsilon=\xi^a \pl_a$ corresponds to the diffeomorphism subalgebra of the gauge symmetries. However, this symmetry acts in a tricky way on the component fields and certain nonlinear field redefinition is needed to make components of $\frepl$ to transform in a usual way. Without such a redefinition, the geometric interpretation of $\psi$ as well as of background fields encoded in $\frepl$ is somewhat unusual. For instance, $\psi$ is naturally a semi-density because the usual $\sqrt{g}$ factor is not present. In order to arrive at the usual "geometrical" interpretation of the fields and their gauge symmetries at least for low-spin fields
one needs to perform a suitable nonlinear field redefinition \cite{Segal:2002gd}.

To conclude, the simple gauge symmetries \eqref{Fpsigauge} encode a rather complicated structure of non-abelian symmetries for higher spin background fields. The partition function 
\begin{align}\label{formalW}
     e^{W_F}&= \int D\bar{\psi} D\psi\, e^{\int i\bar{\psi}\slashed{\pl} \psi + \sum_{s,q} J_{s,q}\tildef_{s,q}}=\int D\bar{\psi} D\psi\, e^{\int \bar{\psi}F \psi}\,,&& W_F=\mathrm{tr}\log F\,,
\end{align}
should be invariant under these symmetries up to an anomaly that is known to show up in even dimensions \cite{Segal:2002gd,Tseytlin:2002gz}. The $\epsilon$-symmetries of \eqref{Fpsigauge} are anomaly-free while the $\alpha,\beta$-symmetries are higher spin extensions of the Weyl symmetry and can be anomalous. The anomaly gives a constructive definition of conformal higher spin theory \cite{Segal:2002gd,Tseytlin:2002gz}. Higher spin sources considered here are easily identified~\cite{Alkalaev:2012ic,Chekmenev:2015kzf} as boundary values of on-shell AdS fields described in section~\bref{sec:ads-free}.

Lastly, when written in terms of $F$ the theory of background fields looks very simple, but it should be remembered that one needs to  choose a vacuum for $F$ to make sense of \eqref{formalW}. For example, the effective action is well-defined when expanded over $i (\gamma\cdot\pl_x)$ and gives the generating functional of correlators of the single-trace operators, as expected. Other choices of vacuum are also possible, e.g. $ (-\pl_x\cdot \pl_x)^k(i\gamma\cdot\pl_x)$ corresponds to so-called higher order singletons --- non-unitary free CFT's originating from $\square^k \slashed \pl\psi=0$.

%%%%%%%%%%%%%%%%%%%%%%%%%%%%%%%%%%%%%%%%%%%%%%%%%%%%%%%%%%%%%
\section{Ambient Space Approach}
\label{sec:ambient}
%%%%%%%%%%%%%%%%%%%%%%%%%%%%%%%%%%%%%%%%%%%%%%%%%%%%%%%%%%%%%

In describing conformal fields as well as their associated background fields it can be very useful to employ the ambient space approach or its generalizations. The underlying idea is to realize the space-time manifold as a quotient of a submanifold of the ambient space where the conformal isometries act naturally and linearly. In the context of (conformal) gravity the generalized version of this procedure (known as Fefferman-Graham ambient metric construction~\cite{FG}) allows one to realize geometrically the Weyl rescalings of the metric. 

%%%%%%%%%%%%%%%%%%%%%%%%%%%%%%%%%%%%%%%%%%%%%%%%%%%%%%%%%%%%%
\subsection{Fermion in Ambient Space}
%%%%%%%%%%%%%%%%%%%%%%%%%%%%%%%%%%%%%%%%%%%%%%%%%%%%%%%%%%%%%
We start with the free Dirac conformal field $\psi(x)$ in Minkowski space, discussed in the preceding section and realize the Minkowski space $\fR^{d-1,1}$ as a space of light-like rays in the ambient space $\fR^{d,2}$ equipped with a pseudo-euclidean metric of signature $(d,2)$. Ambient space Cartesian coordinates are denoted by $X^A$, $A=\{+,-,a\}$, etc.; $\eta_{AB}$ are components of the ambient metric ($\eta^{+-}=\eta^{-+}=1$) and $\cdot$ denotes $o(d,2)$-invariant contraction of ambient indices.

Although the ambient space representation for conformal spinor was introduced already by Dirac~\cite{Dirac:1936fq} (see also~\cite{Marnelius:1978fs}), we need a slightly different formulation, where not all of the constraints are imposed but instead some of them are interpreted as gauge generators. 

It is straightforward to check that the space of configurations for the Minkowski space conformal spinor can be described in terms of the ambient spinor $\Psi(X)$ taking values in the representation $\cS$ of the Clifford algebra generated by ambient $\gamma$-matrices $\Gamma^A$. In order to make $\Psi(X)$ equivalent to a weight-$\Delta$ fermionic field in $d$ dimensions one can impose
\begin{equation}
\label{amb-offshell}
(X\cdot \pl_X+\Delta+\tfrac12)\Psi=0\,, \qquad \qquad \Psi\sim \Psi+X^2\alpha+(X\cdot \Gamma)\beta\,,
\end{equation}
where\footnote{The shift by $\tfrac12$ is due to the fact the $\Psi$ is a spinor and generators of the conformal algebra \eqref{genconf} will compensate for this shift.} the gauge parameters $\alpha$ (resp. $\beta$) also satisfy similar constraints but with $\Delta$ replaced with $\Delta-2$ (resp. $\Delta-1$). Indeed, the constraint and the first equivalence relation imply that we are describing a field on the projectivized hypercone. The second equivalence relation  effectively reduces the representation of Clifford algebra in $(d+2)$-dimensions to that of $d$-dimensions.
To see this, consider the relation at a given point $V^A$ of the hypercone and choose coordinates such that $V^-=1$, $V^+=V^a=0$. Let us represent $\Gamma^+$ and $\Gamma^-$ on Grassmann algebra $\fC[\Gamma^+]$ generated by $\Gamma^+$ and consider $\cS$ as a tensor product $\fC[\Gamma^+]\tensor \cS_0$, where $\cS_0$ is a representation of Clifford algebra generated by $\Gamma^a$. The element in the representation can be written as $\Psi= 1 \tensor \psi_0+\Gamma^+\tensor \psi_1$ where $\psi_{0,1}\in \cS_0$. It is then clear that $\psi_1$ can be set to zero using suitable $\beta$ so that indeed we are dealing with the spinor in $d$ dimensions. Note that with this gauge choice (i.e. elements of the form $1 \tensor \psi_0$, $\psi_{0}\in \cS_0$) $\Gamma^-$ acts trivially.

For $\Delta=\frac{d-1}{2}$ one can impose the  Dirac equation directly in the ambient space.\footnote{More generally one can take $\Delta=\frac{d+1}{2}-\ell$  with $\ell$ positive integer. This choice corresponds to higher-order spinor singletons discussed at the end of the previous section. The generalization of the present discussion to $\ell>1$ can be done in a direct analogy with the higher-order scalar singletons~\cite{Bekaert:2013zya} (see also~\cite{Alkalaev:2014qpa,Bekaert:2017bpy}). Representation theoretical study of these fields can be found in~\cite{Basile:2014wua}.} Indeed,
one can extend the above system by the following equations:
\begin{equation}
\label{ambeom}
    \pl_X\cdot\pl_X\Psi=0\,, \qquad\qquad     \Gamma \cdot \pl_X\Psi=0\,,
\end{equation}
where the first one is a consequence of the second. It is easy to check that the full system is indeed consistent for $\Delta=\frac{d-1}{2}$ provided $\alpha$, $\beta$ are subject to the analogous constraints. It can be shown that altogether \eqref{amb-offshell}, \eqref{ambeom} imply the Dirac equation in $d$-dimensions. A simple though not rigorous argument is to use gauge freedom to choose $\Psi$ to be $X^-$-independent and such that $\Gamma^-\Psi=0$ (strictly speaking this can be done only at a point). With this choice the equations reproduce the Dirac equation and its consequence, the Klein-Gordon equation in Minkowski space. % 

We see that the constrained system \eqref{freeconstraints}, whose relation to conformal symmetry is not manifest, can be replaced by the constrained system \eqref{amb-offshell}, \eqref{ambeom}. This system is determined by the five operators\footnote{Note that the constant in the third operator differs from that in~\eqref{amb-offshell}, the origin of this shift is that in~\eqref{amb-offshell} $X^2$ and $X\cdot \Gamma$ are gauged rather than imposed, see e.g.~\cite{Grigoriev:2011gp,Bekaert:2012vt}.}
\begin{equation}\label{freeosp}
X^2\,,\qquad \d_X^2\,,\qquad (X\cdot\d_X+\tfrac{d+2}{2})\,, \qquad \Gamma\cdot X\,, \qquad \Gamma\cdot\d_X\,.
\end{equation}
The five \eqref{freeosp} form a representation of $osp(1|2)$-superalgebra on functions in $X$.
In this representation the $osp(1|2)$ generators \eqref{freeosp} commute with the conformal algebra $o(d,2)$ represented in a standard way:
\begin{equation}\label{genconf}
J_{AB}=X_A\dl{X^B}-X_B\dl{X^A}+\tfrac1{4}(\Gamma_A\Gamma_B-\Gamma_B\Gamma_A)\,,
\end{equation}
which makes conformal symmetry manifest. In this representation $o(d,2)$ and $osp(1|2)$ form a reductive dual pair in the sense of Howe~\cite{Howe,Howe1}. In particular, the conformal group acts on equivalence classes determined by~\eqref{amb-offshell}
as well as on solutions to \eqref{amb-offshell}, \eqref{ambeom}. In this way one can easily re-derive the conformal transformations of the fermionic operator $\psi(x)$ in $d$ dimensions.

The description just given is nothing but a usual first-quantized description in which $\Psi$ is the wave function of the first-quantized constrained Hamiltonian system describing a particle with spin $1/2$. Once we have a manifestly $o(d,2)$-invariant description of the free fermion in terms of the $osp(1|2)$ constraints, the next step is to add higher spin background fields, which is similar to going from \eqref{freeconstraints} to \eqref{constraints}.

%%%%%%%%%%%%%%%%%%%%%%%%%%%%%%%%%%%%%%%%%%%%%%%%%%%%%%%%%%%%%
\subsection{Background Fields in Ambient Space}
%%%%%%%%%%%%%%%%%%%%%%%%%%%%%%%%%%%%%%%%%%%%%%%%%%%%%%%%%%%%%
Given a first-quantized constrained system, one can systematically derive, in general nonlinear, equations and gauge symmetries for background fields. Such procedure was discussed in the context of BRST formulation of string theory in~\cite{Horowitz:1986dta}. A general approach suitable in the present context was developed in~\cite{Grigoriev:2006tt}, see also \cite{Bekaert:2013zya}. 

Now we briefly recall the basic ideas. We restrict ourselves to the description at the level of equations of motions. At this level there is no need to explicitly introduce representation space of the quantum system and it is useful to employ the language of the star-product so that instead of quantum operators we work with phase space functions and use star-product instead of operator multiplication. Moreover, we do not impose reality conditions and hence work with complexified fields.

Suppose we are given the first class constrained system with constraints $F_i$ which are elements of the (graded) star-product algebra.  
The consistency condition reads as\footnote{The structures we discuss below can be systematically extracted from the BRST operator encoding these constraints, see~\cite{Grigoriev:2006tt,Bekaert:2013zya} for more detail. To simplify the discussion we prefer not to go into the details of the BRST formulation.}
\begin{equation}\label{ext1}
[F_i,F_j]_\star=U^k_{ij} \star F_k\,,
\end{equation}
where, as before, $\qcommut{\,}{\,}$ denotes $\star$ super-commutator determined by the Grassmann degree in the $\star$-product algebra. Here and in the sequel we assume that for a given $F_i$ its Grassmann degree $\p{i}$ is set. In particular, in the case where $U^k_{ij}$ are structure constants of a Lie superalgebra,  $\p{F_i}$ (as an element of the $\star$-product algebra) must coincide with the degree $\p{i}$ of a basis element $e_i$ of the Lie superalgebra, to which $F_i$ is associated, see~\cite{Alkalaev:2014nsa} for more detail.

If we identify functions $F_i$ as generating functions for background fields, \eqref{ext1} can be interpreted as equations of motion. Moreover, natural equivalence transformations for the constraints 
\begin{equation}\label{ext2}
\delta F_i=\lambda^j_i \star F_j+\qcommut{\epsilon}{F_i}\,,
\end{equation}
can be interpreted as gauge transformations for $F_i$. Note that in general $U_{ij}^k$ are also transformed, see e.g. \cite{Alkalaev:2014nsa} for more detail. 

The transformations with $\lambda^i_j$  correspond to infinitesimal redefinitions of the constraints (at classical level such symmetries preserve the constraint surface; at the quantum level they preserve the physical subspace). In this way one arrives at the gauge theory of background fields associated with a given constrained system. From this perspective, the constrained system itself (i.e. when $F_i$, $U_{ij}^k$ are concrete phase space functions) is just a fixed solution (e.g. vacuum solution) to the theory of background fields.

One can also think of the space of all reasonable solutions quotiented by the above gauge symmetry as the moduli space of constrained systems with a given phase space. It is important to note that one and the same theory of background fields may have various inequivalent vacua. Examples of such vacua have already been given at the end of section \bref{sec:fermion-mink}: $F=i(\gamma\cdot p) (-p^2)^k$ correspond to different free CFT's whose off-shell field content is given by the same $\psi(x)$. Note that for different $k$ the spectrum of single-trace operators and hence the set of background fields are different, so that the vacua are clearly non-inequivalent.

Background fields for Dirac fermion on Minkowski space discussed in section~\bref{sec:fermion-mink} is a simple example of the general pattern. Because the ambient $osp(1|2)$ constrained system from the previous section is equivalent to the Minkowski space one, we can equivalently use it to derive a more convenient ambient form for the associated theory of background fields.\footnote{This trick was originally used in~\cite{Grigoriev:2006tt} in the context of background fields for AdS scalar.} This still has a form~\eqref{ext1}, \eqref{ext2} where $F_i$ are associated to the $osp(1|2)$ generators and are functions on the ambient phase space with coordinates $X^A$, $P_B$, $\Mph_C$. The $\star$-product is determined by
\begin{equation}
\qcommut{X^A}{P_B}=\delta^A_B\,, \qquad\qquad 
\qcommut{\Mph_A}{\Mph_B}=2\eta_{AB}\,,
\end{equation}
where the Grassmann degree in the $\star$-product algebra is introduced according to $\p{\Mph}=1,\,\,\p{X}=\p{P}=0$.

Let us recall that in the minimal construction we are concerned now we require $\p{F_i}=\p{i}$, where $\p{i}$ denotes Grassmann degree of the $osp(1|2)$ generator to which $F_i$ is associated. In particular, $F_i$ associated to odd generators of $osp(1|2)$ contain only odd powers of $\Mph$ and hence, as we are going to see, the system describes the minimal multiplet of hook-shape fields (for $d$-odd this gives a multiplicity free set of all irreducible hook-shape tensors; for $d$ even these are fields associated to all Young diagrams of odd height).\footnote{
A possible way to describe nonminimal multiplets (e.g. to have background fields of even height in the case of even $d$) is to extend the algebra with an extra Clifford generator $\kkk$ satisfying
\begin{equation}
\kkk\star\kkk=1\,,  \qquad  \kkk\star\Mph_A+\kkk\star\Mph_A=0\,, \qquad  \p{\kkk}=1\,,
\end{equation}
and not entering generators for $osp(1|2)$ and $o(d,2)$-algebras.
With this extension the homogeneity in $\Mph$ is not restricted despite the condition $\p{F_i}=\p{i}$.
This trick is based on the standard fact that Clifford algebra can be realized as an even subalgebra of the Clifford algebra in one dimension higher.}

For system~\eqref{ext1}, \eqref{ext2} to describe background fields for the Dirac fermion we need to restrict ourselves to $F_i$ which are close to the vacuum solution ($i=\{++,0,--,+,-\}$)
\begin{equation}
\label{osp12-vacuum}
F^0_{++}=\half P^2\,, \qquad F^0_0=X\cdot P\,,  \qquad    
F^0_{--}=\half X^2\,, \qquad F^0_+ =\Mph\cdot P\,, \qquad 
F^0_-=\Mph\cdot X\,.
\end{equation}
Moreover, we need to consider the system in the vicinity of the hypercone $X^2=0$ in the ambient space. With these precautions one can indeed prove, see below, that this gives an equivalent description of the background fields of section \bref{sec:fermion-mink}. Somewhat analogous ambient space approach to background fields for the Dirac fermion was considered in~\cite{Bonezzi:2015bza} from a slightly different perspective, see also~\cite{Bars:2000qm,Bonezzi:2010jr,Bonezzi:2014nua} for earlier related works.

To conclude, the manifestly $o(d,2)$-invariant formulation of background fields for Dirac fermion has the form of the $osp(1|2)$ system of constraints. Once formulated in ambient space, any system can, at least formally, be considered in the vicinity of the hyperboloid $X^2=-1$ to give a theory in $AdS_{d+1}$.\footnote{Although the idea to use ambient space to go from bulk to boundary and back is known starting from the Fefferman-Graham ambient metric construction the framework we are going to use was developed in \cite{Bekaert:2012vt,Bekaert:2013zya,Chekmenev:2015kzf} and employed in a similar context recently in~\cite{Bekaert:2017bpy}.} Before doing that, let us review what is the expected free field content of the dual theory and how to describe it in the ambient space.

%%%%%%%%%%%%%%%%%%%%%%%%%%%%%%%%%%%%%%%%%%%%%%%%%%%%%%%%%%%%%
\section{Type-B Theory at Free Level}
\label{sec:ads-free}
%%%%%%%%%%%%%%%%%%%%%%%%%%%%%%%%%%%%%%%%%%%%%%%%%%%%%%%%%%%%%
We begin with developing a compact formulation of unitary gauge fields in $AdS_{d+1}$, which are tensors  whose symmetry is described by Young diagrams of the hook shape:\footnote{Indices $\mu,\nu,...$ are the indices of $AdS_{d+1}$ in some local coordinates. Field $\Phi^{\mu_1...\mu_s,\nu_1...\nu_q}$ is symmetric in the $\mu$'s and anti-symmetric in the $\nu$'s. The Young symmetry implies $\Phi^{\mu_1...\mu_s,\mu_{s+1}\nu_2...\nu_q}+\text{permutations of $\mu$'s}\equiv0$.} 
\begin{align}
    \Phi_{\mu_1...\mu_s,\nu_1...\nu_q}(x)\sim
    \parbox{60pt}{{\bep(60,50)\unitlength=0.38mm%
    \put(0,40){\RectARowUp{6}{$s$}}%
    \put(0,0){\RectT{1}{4}{\TextCenter{$q$}}}\eep}}
\end{align}
These fields are in one-to-one with the sources from the previous sections in the sense that the later can be seen as the leading boundary values of the former~\cite{Alkalaev:2012ic,Chekmenev:2015kzf}. In the next section we are going to extract these fields from the non-linear system.

As different from totally-symmetric fields that constitute the spectrum of the Type-A higher spin gravity that is dual to the free boson CFT, the spectrum of the Type-B theory contains fields of the symmetry type depicted above. Gauge fields of mixed-symmetry are trickier and reveal some features not present in the totally-symmetric case \cite{Metsaev:1997nj}.

Not going into the detail, we simply recall that the equations of motion and gauge symmetries are~\cite{Metsaev:1997nj,Alkalaev:2009vm}:
\begin{align}
    (-\nabla^2 +M_{s,q}^2)\Phi_{\mu_1...\mu_s,\nu_1...\nu_q}(x)&=0\,,\\
    \delta \Phi_{\mu_1...\mu_s,\nu_1...\nu_q}(x)&= \nabla_{\mu_1} \xi_{\mu_2...\mu_s,\nu_1...\nu_q}(x)+ \text{permutations}\,.
\end{align}
It is assumed that both the field and the gauge parameter are in the transverse-traceless gauge, i.e. all traces and divergences vanish:
\begin{align}
    \nabla^\lambda \Phi_{\lambda \mu_2...\mu_s,\nu_1...\nu_q}(x)&=0\,, &
    \nabla^\lambda \Phi_{\mu_1...\mu_s,\lambda \nu_2...\nu_q}(x)&=0\,,\\
    g^{\lambda\rho}\Phi_{\lambda\rho \mu_3...\mu_s,\nu_1...\nu_q}(x)&=0\,, &
    g^{\lambda\rho}\Phi_{\lambda \mu_2...\mu_s,\rho \nu_2...\nu_q}(x)&=0\,,
\end{align}
and similarly for the gauge parameter. 
As a result, the gauge parameter has to obey a similar on-shell condition. The value of the mass parameter is fixed by gauge invariance to be $M_{s,q}^2= \Delta(\Delta-d)-s-q$ \cite{Metsaev:1997nj}, where the cosmological constant was set to one and $\Delta=d+s-2$ for the duals of conserved tensors. It is worth mentioning that not all of the fields in the hypothetical Type-B theory are gauge fields. Those with $s=1$ and $q=1,2,...$ are massive and $M^2_{s=1,q}=-(d-1)-q-1$. There is also a scalar field $s=0,q=0$ with $M^2_{s=0,q=0}=-(d-1)$. 

%%%%%%%%%%%%%%%%%%%%%%%%%%%%%%%%%%%%%%%%%%%%%%%%%%%%%%%%%%%%%
\subsection{Ambient Space Description}
%%%%%%%%%%%%%%%%%%%%%%%%%%%%%%%%%%%%%%%%%%%%%%%%%%%%%%%%%%%%%

It is convenient to describe these fields in terms of the same ambient space $\fR^{d,2}$ with coordinates $X^A$ and such that $AdS_{d+1}$ space can be (locally) identified with the hyperboloid $X\cdot X=-1$. To pack the fields into generating functions it is also convenient to introduce commuting  variables $P_A$  to contract the indices associated to the upper row of the Young diagram and anticommuting variables $\Mph_A$ to contract the indices associated to the first column (save for the upper cell) of the Young diagram, the generating function being
\begin{equation}
\label{symm-base}
    f_S(X,P,\Mph)=\sum_{\substack{s\geq 1,q\geq 0\\ s=0,q=0} }f^{A_1\ldots A_s; B_{1}\ldots B_{q}}_S\,P_{A_1}\ldots P_{A_s}\Mph_{B_1}\ldots\Mph_{B_{q}}\,,
\end{equation}
which encodes all fields of this type simultaneously.
The usage of (anti)commuting variables ensures that the field is symmetric in the first group of indices and is antisymmetric in the second one. The Young symmetry conditions can be compactly expressed in terms of the above generating functions as $(P\cdot\d_\Mph) f_S=0$.

The ambient space field $f_S$ introduced above can be subject to the constraints which effectively reduce it to the collection of tensor fields of the same Young shape defined on the hyperboloid $X^2=-1$. The constraints are:
\begin{equation}
(X\cdot\d_P)f_S=0\,, \quad
(X\cdot\d_\mph)f_S=0\,, \quad 
(P\cdot\d_\Mph)f_S=0\,, \quad 
(X\cdot \d_X-P\cdot \d_P+w)f_S=0\,,
\end{equation}
where $w$ is a number which we leave free at the moment. The ambient field $f_S$ can also be subject to $o(d,2)$-invariant extra constraints
\begin{equation}
(\d_X\cdot\d_X)f_S=(\d_X\cdot\d_P)f_S=(\d_P\cdot\d_P)f_S=(\d_X\cdot\d_\mph)f_S=(\d_P\cdot\d_\mph)f_S=0\,,
\end{equation}
which imply that the component fields are irreducible tensors subject to the usual 2-nd order differential equations. In fact, for generic weight $w$ the system describes massive fields on $AdS_{d+1}$. 

If one takes $w=2$,
the above system possesses gauge symmetries
$\delta f_S=(P\cdot\d_X) \xi_S$, where $\xi_S$ is a gauge parameter subject to the analogous constraints. Let us summarize the constraints on $f_S$ and $\xi_S$ and the gauge transformations:
\begin{equation}
\label{eoms+traces}
\begin{gathered}
(\d_X\cdot\d_X)f_S=(\d_X\cdot\d_P)f_S=(\d_P\cdot\d_P)f_S=(\d_X\cdot\d_\mph)f_S=(\d_P\cdot\d_\mph)f_S=0\,,\\
(\d_X\cdot\d_X)\xi_S=(\d_X\cdot\d_P)\xi_S=(\d_P\cdot\d_P)\xi_S=(\d_X\cdot\d_\mph)\xi_S=(\d_P\cdot\d_\mph)\xi_S=0\,,
\end{gathered}
\end{equation}
and 
\besubeqs
\label{amb-hook}
\begin{align}
   (X\cdot \d_\mph) f_S =(X\cdot \d_P) f_S=(P\cdot \d_P-X\cdot \d_X-2)f_S=( P\cdot\d_\mph)f_S=0\,, \\
   (X\cdot \d_\mph) \xi_S = (X\cdot \d_P) \xi_S=(P\cdot \d_P-X\cdot \d_X)\xi_S=(P\cdot\d_\mph)\xi_S=0\,,\\
    \delta f_S=(P\cdot\d_X) \xi_S\,.\qquad\qquad\qquad\qquad\qquad\qquad
\end{align}
\esubeqs
Specializing the ambient formulation~\cite{Alkalaev:2009vm} of generic AdS fields to massless unitary gauge fields of hook-type, and rewriting the formulation of~\cite{Alkalaev:2009vm} in terms of fermionic  $\mph_A$ in place of of the bosonic oscillators one precisely arrives at~\eqref{eoms+traces} and \eqref{amb-hook}.

Note that all the constraints and the gauge generator clearly form a subalgebra of $osp(4|2)$ (strictly speaking some constant terms are to be added). In this representation this algebra commutes with $o(d,2)$ that acts by
\begin{equation}
    J_{AB}=X_A\dl{X^B}+P_A\dl{P^B}+\mph_A\dl{\mph^B}-(A\rightleftarrows B)\,.
\end{equation}
More specifically, if we restrict ourselves to polynomials in $X,P,\mph$ we get the standard setting of the Howe duality.

To complete the discussion of the hook-type fields let us present an alternative formulation, which, as we are going to see, arises as a linearization of the Type-B non-linear theory. In terms of the generating functions $f_+(X,P,\mph)$, $\xi(X,P,\mph)$ for fields and parameters the equations of motion and gauge symmetries take the form:
\besubeqs
\label{linearized-hook}
\begin{align}
\label{const-1}
   (X\cdot \d_\mph+\mph\cdot \d_P)f_+ =(X\cdot \d_P) \,f_+= (P\cdot \d_P-X\cdot \d_X-1)f_+=0\,, \\
   (X\cdot \d_\mph+\mph\cdot \d_P)\xi = (X\cdot \d_P) \xi=(P\cdot \d_P-X\cdot \d_X)\xi=0\,,\\
\label{const-13-gs}
    \delta f_+=(P\cdot \d_\mph-\mph \cdot \d_X)\xi\,,\qquad\qquad\qquad\qquad
\end{align}
\esubeqs
and
\begin{equation}
\label{linearized-hook-eoms+traces}
\begin{gathered}
(\d_X\cdot\d_X)f_+=(\d_X\cdot\d_P)f_+=(\d_P\cdot\d_P)f_+=(\d_X\cdot\d_\mph)f_+=(\d_P\cdot\d_\mph)f_+=0\,,\\
(\d_X\cdot\d_X)\xi=(\d_X\cdot\d_P)\xi=(\d_P\cdot\d_P)\xi=(\d_X\cdot\d_\mph)\xi=(\d_P\cdot\d_\mph)\xi=0\,.
\end{gathered}
\end{equation}
Note that neither fields nor gauge parameters are irreducible tensors in this formulation. 
In Appendix~\bref{sec:symmbase} we show that
this formulation is equivalent to~\eqref{amb-hook}, \eqref{eoms+traces} through partial gauge fixing and field redefinition. Also, c.f. \eqref{const-13-gs} and \eqref{linsym}.

In what follows, we mostly make use of minimal subset of the above fields which still admits nonlinear extension. The subset is obtained by setting to zero all $f_+$ that are of even homogeneity degree in $\Mph$ (in the formulation in terms of $f_S$ one sets to zero even degree fields). Note that for $d$-odd the subset is multiplicity free in the sense that each irreducible massless field contained in the subset enters only once and moreover all inequivalent massless fields of hook shape are present.

In what follows it is also convenient to identify the off-shell version of the above linear theory. It is obtained by dropping constraints \eqref{eoms+traces} or \eqref{linearized-hook-eoms+traces}. 

%%%%%%%%%%%%%%%%%%%%%%%%%%%%%%%%%%%%%%%%%%%%%%%%%%%%%%%%%%%%%
\section{Type-B Theory}
\label{sec:typeb}
%%%%%%%%%%%%%%%%%%%%%%%%%%%%%%%%%%%%%%%%%%%%%%%%%%%%%%%%%%%%%
In the section we propose a non-linear formally consistent gauge invariant system that reproduces Type-B theory at the free level and also reproduces all the structures determining the formal non-linear deformation. The theory's boundary values are the background fields $\tildef^{s,q}$ introduced in section \bref{sec:fermion-mink} together with the symmetries thereof and the latter, as we know, completely fix the effective action $W[\tildef]$.

Let us consider again the constrained system \eqref{ext1}, \eqref{ext2} defined in the ambient space in the vicinity of the hypercone $X^2=0$ and assume that fields are close to their vacuum values~\eqref{osp12-vacuum} and hence $U_{ij}^k$ are close to the $osp(1|2)$ structure constants $C_{ij}^k$. As we already discussed, this system describes the same background fields $\tildef^{s,q}$ for Dirac fermion in $d$-dimensions realized in the ambient space.

The idea is to uplift this system from the conformal $d$-dimensional space (boundary) to $(d+1)$-dimensional AdS space (bulk) in such a way that the boundary values of the fields of the bulk theory coincide with the background fields $\tildef^{s,q}$ on the boundary, while preserving all the symmetries of the latter. The ambient formulation suggests an easy way to do this by considering \eqref{ext1}, \eqref{ext2} to be defined in the vicinity of the hyperboloid $X^2=-1$. This is justified by the fact that within the ambient framework it is known~\cite{Bekaert:2012vt,Bekaert:2013zya} that the  passage from bulk fields to their boundary values amounts to simply considering the same ambient system in the vicinity of $X^2=0$ rather than $X^2=-1$.

To see that we are on a right track, let us temporarily disregard gauge transformations \eqref{ext2} with parameters $\lambda$, i.e. set $\lambda^i_j$ to zero in~\eqref{ext2}. The remaining gauge transformations with parameter $\epsilon$ preserve the structure constants so that we can assume $U_{ij}^k=C_{ij}^k$. Then it turns out that the linearization of  \eqref{ext1}, \eqref{ext2} around the vacuum solution \eqref{osp12-vacuum} followed by partial gauge fixing and solving some of the equations results in~\eqref{amb-hook}, i.e. reproduce the off-shell Type-B theory at the free level. The detailed proof of this statement is given in the next section, where we introduce more powerful formalism to handle the system.

If one reinstates the gauge transformations with parameters $\lambda^i_j$ the interpretation of the system drastically depends on the choice of the functional class for fields. To see this, we restrict ourselves to field configurations that are close to the vacuum solution~\eqref{osp12-vacuum}.\footnote{The interpretation also depends on the choice of vacuum. Let us remind that this is true even on the CFT side, where the situation is much more clear: different choices of $F$ in \eqref{FHsystem} (or $F_i$ in \eqref{ext1}) correspond to different CFT's. For instance, if in \eqref{FHsystem} we choose $F=1+...$ instead of $F=i (\gamma\cdot p)+...$ then the CFT is empty. Let us note that our understanding of the functional class problems on the CFT side is also somewhat illusory: the effective action $W[F]=\mathrm{tr}\log F$ resulting from the formal path integral is just a symbol that can be made sense of in special situations only. The functional class issues we are facing should be related to the general problem of how to relate Hilbert spaces on the two sides of AdS/CFT.} Then on one hand, if $F_i$ are smooth functions defined in a vicinity of the hyperboloid this gauge symmetries can be used to gauge away all $F_i$ because $F_{--}=-1+\ldots$ on the hyperboloid (cf. \eqref{osp12-vacuum}) and hence can be inverted. On the other hand, if $F_i$ are polynomials in $X$ this gauge symmetry can be used to make all $F_i$ totally traceless, i.e. satisfying~\eqref{eoms+traces} as required for the free on-shell Type-B theory. However, this functional class is not suitable for genuine field theory (fields are not polynomials). 

Therefore, the problem boils down to identifying the right functional class for fluctuations of $F_i$. It turns out that there is a consistent choice of functional class  such that the linearization of \eqref{ext1}, \eqref{ext2}
 is indeed equivalent to \eqref{amb-hook}, \eqref{eoms+traces}, i.e. reproduces the on-shell Type-B theory at the free level.
 This functional class is a straightforward generalization of the one, which was employed in the analogous construction~\cite{Bekaert:2017bpy} for the Type-A.
 
 In order to study the system further and to introduce the functional class it is useful to reformulate it using the so-called parent formalism. For theories of background fields the construction was originally proposed in~\cite{Grigoriev:2006tt} while the general approach was developed in~\cite{\BGST,Barnich:2006pc,Barnich:2010sw,Grigoriev:2010ic}. In this approach a given system is reformulated as an AKSZ-type sigma-model~\cite{Alexandrov:1995kv} whose target space is the jet-space BRST complex of the gauge theory under considerations.
 
 The geometric idea of the parent reformulation in the present context is to consider the theory as defined on the formal ambient space with coordinates $Y^A$ and then identify this space as a fiber of a fiber-bundle over the genuine space-time manifold ($AdS_{d+1}$ in our case). In so doing one gets a collection of systems parameterized by space-time coordinates. The equivalence is then maintained by introducing an extra field, the connection one-form $A$ with values in the $\star$-product algebra and to require this connection to be flat and all the fields to be covariantly constant. As a byproduct, one arrives at the reformulation of the original ambient space theory  in terms of fields explicitly defined on AdS space.

This being said, let us construct parent formulation of the ambient space system~\eqref{ext1}-\eqref{ext2}. The target space is the (super) Weyl algebra of $(d+2)$ canonical bosonic pairs $Y^A$, $P_B$ and $(d+2)$ fermionic $\Mph^A$. In practice, we deal with the star-product algebra of functions in $Y^A$, $P_B$ and $\Mph^A$:
\begin{align}
    (f\star g)(Y,P,\Mph)&= f(Y,P,\Mph) \exp\left[\frac{\overleftarrow{\pl}}{\pl Y^A} \frac{\overrightarrow{\pl}}{\pl P_A}-\frac{\overleftarrow{\pl}}{\pl P_A} \frac{\overrightarrow{\pl}}{\pl Y^A}+\frac{\overleftarrow{\pl}}{\pl \Mph^A} \frac{\overrightarrow{\pl}}{\pl \Mph_A}\right]g(Y,P,\Mph)\,.
\end{align}
The field content consists of five zero-forms $F_i$ and one one-form connection $A$:
\begin{align}
    A&=dX^{\uB} A_{\uB} (X|Y,P,\Mph)\,, &&F_i=F_i(X|Y,P,\Mph)\,,
\end{align}
where $X^{\uB}$ denotes local coordinates on the ambient space. The Grassmann degree naturally extends to the new variables and fields and is determined by:
\begin{equation}
\label{g-degree}
\p{\Mph}=\p{dX}=1\,, \qquad \p{Y}=\p{P}=\p{X}=0\,, \qquad \p{F_i}=\p{i},\quad  \p{A}=1\,.
\end{equation}

%%%%%%%%%%%%%%%%%%%%%%%%%%%%%%%%%%%%%%%%%%%%%%%%%%%%%%%%%%%%%
\subsection{Off-shell System}\label{sec:off-shell}
%%%%%%%%%%%%%%%%%%%%%%%%%%%%%%%%%%%%%%%%%%%%%%%%%%%%%%%%%%%%%

We first consider the off-shell version of the parent system. It is constructed out of the ambient space system \eqref{ext1}, \eqref{ext2} where the $\lambda$-gauge symmetries related to the redefinition of the constraints have been disregarded. In this case one can assume $U^k_{ij}=C^k_{ij}$ so that the equations of motion and the gauge symmetries read as:\\
\besubeqs\label{Bsystem}\begin{align}
    dA&=\tfrac12 [A,A]_\star\,, & \delta A&=d\xi-[A,\xi]_\star\,,\\
    dF_i&= [A,F_i]_\star\,, & \delta F_i&= [\xi,F_i]_\star \,,
    \label{cov-const-F}\\
    [F_i,F_j]_{\star}&=C_{ij}^k F_k\,,
\end{align}\esubeqs
where $d$ is the de Rham differential on the spacetime manifold and $\qcommut{}{}$ is a super-commutator defined with respect to Grassmann degree~\eqref{g-degree}. Note that now $X^{\uA}$\,-variables commute with all the other and merely serve as parameters. 

The equivalence with the original system~\eqref{ext1}-\eqref{ext2} can be seen by requiring $A$ to be ``sufficiently close'' to the
vacuum solution
\begin{align}
A^0=dX^AP_A\,.
\end{align}
With $A=A^0$ the first equation in \eqref{cov-const-F} implies $\dl{X^A}F_i=\dl{Y^A}F_i$, so that the equivalence is straightforward. Furthermore, the linearized gauge transformations for $A$ involve $dX^A\dl{Y^A} \xi$, which implies that at least linearized fluctuations can be gauged away thanks to the Poincare Lemma for the formal de Rham differential $dX^A\dl{Y^A}$. In other words, for $A$ sufficiently close to $A^0$ one can assume that the gauge $A=A^0$ is reachable. 

In the gauge $A=A^0$ it is easy to identify a parent version of the vacuum solution
~\eqref{osp12-vacuum}
\begin{equation}
\label{vacuum-ambient}
\begin{gathered}
A=A^0\,,      \qquad F^0_-= (Y^A+X^A)\mph_A\,, \qquad
    F^0_+=P^A\mph_A\,, 
\\
    F^0_{--}=\frac12 (Y+X)\cdot(Y+X)\,, \qquad
    F^0_{++}=\frac12 P\cdot P\,,  \qquad
    F^0_0= (Y+X)\cdot P\,.
\end{gathered}
\end{equation}
Nevertheless, it is very useful not to gauge-away $A$ in order to be able to work in any coordinates on the base space and allow for rather general local transformations in the fiber. Note that in the classical ($\star$-commutator replaced by the Poisson bracket) limit $A$ can be taken linear in $P_A$ and is nothing but the gauge field associated to ambient space diffeomorphisms.

If we consider~\eqref{Bsystem} in the vicinity of the hyperboloid and interpret it as a local field theory defined on the hyperboloid $X^2=-1$, it is convenient to pull-back the system to $X^2=-1$ in order to explicitly work with fields defined in terms of generic coordinates on the hyperboloid. With a suitable choice of a local frame the adapted version of the vacuum solution~\eqref{vacuum-ambient} reads as\footnote{Here $x^\mu$ are some intrinsic coordinates on the hyperboloid and $dx^\mu$ are the associated differentials.}
\begin{equation}
\label{vacuum-parent}
\begin{gathered}
A^0=dx^\mu\omega_{\mu}{}_{A}{}^B T\fud{A}{B}\,,\qquad\qquad T^{AB}=-(Y^A+V^A)\cdot P^B+\frac{1}{4}\mph^A\mph^B -(A\rightleftarrows B)\,,\\
F^0_-= (Y^A+V^A)\mph_A\,,\qquad\qquad
    F^0_+=P^A\mph_A\,,
\\
    F^0_{--}=\frac12 (Y+V)\cdot(Y+V)\,,\qquad
    F^0_{++}=\frac12 P\cdot P\,, \qquad
    F^0_0= (Y+V)\cdot P\,,
\end{gathered}
\end{equation}
where $T^{AB}$ are the $o(d,2)$ generators, the compensator field $V^A$ is taken constant, $V\cdot V=-1$, and $\omega_A^B$ is a flat $o(d,2)$-connection one-form such that
$\nabla_\mu V^A$ has maximal rank. Compensator field $V^A$ has a clear meaning of the original Cartesian coordinate $X^A$ on the ambient space which was set to constant by a local $o(d,2)$-transformations (recall that the theory is defined on the hyperboloid so that $V^2=-1$). Note that such a vacuum solution exists even if a spacetime is isometric to $AdS_{d+1}$ only locally.

In what follows we refer to system~\eqref{Bsystem} defined on $(d+1)$-dimensional space as to {\it off-shell system}. Unless otherwise specified  it is assumed that the theory is understood around the vacuum where $F=F^0_i$, with $F^0_i$ defined in~\eqref{vacuum-parent}.

\paragraph{Linearization.}
The linearized equations are obtained from \eqref{Bsystem} by replacing $A\rightarrow A^0+a$ and $F_i\rightarrow F_i^0+f_i$ and picking the terms linear in $a$ and $f_i$:%
\besubeqs
\label{linearized-p}
\begin{align}
    D_0a&=0\,, & \delta a&=D_0\xi\,,\\
    D_0f_i&= [a,F_i^0]_\star\,, & \delta f_i&= [\xi,F_i^0]_\star\,, \\
    [F^0_i,f_j]_{\star}-(ij)-C_{ij}^k f_k&=0\,,
\end{align}
\esubeqs%
where $A^0$, $F^0_i$ is a vacuum solution and $D_0\bullet \equiv d\bullet -[A^0,\bullet]$ is the background covariant derivative. In our case, the vacuum is \eqref{vacuum-parent}.

It is easy to check that the system \eqref{linearized-p} can be written in the BRST first quantized form 
\begin{equation}
\label{BRST-form}
\Omega \Phi=0\,, \quad \delta \Phi=\Omega \Xi\,, \qquad \gh{\Phi}=1\,, \quad \gh{\Xi}=0\,,
\end{equation}
introducing ghost variables $c^i$ and using the BRST operator
\begin{equation}
\label{BRST-op}
\Omega=D_0+c^i\qcommut{F^0_i}{\bullet}-(-1)^{\p{i}(\p{j}+1)}\half c^i c^j C_{ij}^k \dl{c^k}  \,.
\end{equation}

Then working in the local frame such that $V^A=(1,0,\ldots,0)$ and using notations $\bar y,y^m$ for the respective components of $Y^A$ and similarly for $P_A,\Mph_A$ let us decompose the representation space (i.e. the space where $\Phi$ takes values) according to the following degree
\begin{equation}
\deg{c^{++}}=\deg{c^{+}}=\deg{dx^\mu}=1\,.
\end{equation}
Accordingly, the BRST operator $\Omega$-decomposes into homogeneous pieces as 
\begin{equation}\notag
\Omega=\Omega_{0}+\Omega_{1}\,, \qquad \Omega_{0}=c^0\qcommut{(Y+V)\cdot P}{\bullet}+c^{-}\qcommut{(Y+V)\cdot\mph}{\bullet}+c^{--}\qcommut{\tfrac12(Y+V)^2}{\bullet}+\text{ghosts}\,.
\end{equation}
The operator $\Omega_{0}$ is algebraic and it is known~\cite{Barnich:2004cr} that the BRST system can be equivalently reduced to the one
whose representation space is identified with the cohomology of $\Omega_{0}$. 

To compute the cohomology of $\Omega_0$ we use another degree in the representation space:
\begin{equation}
\deg^\prime{\bar y}=\deg^\prime{\bar p}=\deg^\prime{\bar \mph}=1\,,
\end{equation}
and hence $\Omega_0$ can be decomposed as
\begin{equation}
\Omega_0=\Delta_{-1}+\Delta_0\,, \qquad \Delta_{-1}=-c^{0}\dl{\bar y}+c^{--}\dl{\bar p}+2c^{-}\dl{\bar \mph}\,,
\end{equation}
with $\deg^\prime{\Delta_i}=i$. Cohomology of $\Omega_0$ can be computed by first computing cohomology of $\Delta_{-1}$. In its turn, the cohomology of $\Delta_{-1}$
is clearly given by $c^0,c^{--},c^-,\bar y,\bar p,\bar\mph$-independent elements and hence representatives of $\Omega_0$\,-cohomology can be taken to be $c^0,c^{--},c^-$-independent. The system is then  reduced to (the reduction amounts to just restricting $\Omega_1$ to act on the subspace of the representatives because $\Omega_1$ preserves this subspace):
\begin{equation}
\begin{gathered}
\Omega^{\prime}=D_0+c^{++}\qcommut{\half P^2}{\bullet}+c^+\qcommut{P\cdot \mph}{\bullet}-2 c^+ c^+ \dl{c^{++}}\,,\\ \Phi^{\prime}=\Phi^{\prime}(x,dx^\mu,Y,P,\mph,c^{++},c^{+})\,, \qquad \Omega_0\Phi^{\prime}=0\,. 
\end{gathered}
\end{equation}

It is useful to perform one more reduction by going to cohomology of the term
$c^+ c^+ \dl{c^{++}}$ entering $\Omega^{\prime}$. The cohomology can be explicitly realized as the quotient space of $c^{++}$-independent elements by those proportional to $(c^+)^2$ and it is useful to work in terms of representatives that are at most linear in $c^+$. Then the reduced system is given by 
\begin{equation}
\begin{gathered}
\label{omega-red}
\Omega^{red}=D_0+c^+\qcommut{P\cdot \mph}{\bullet}\,,\\ \Phi^{red}=\Phi^{red}(x,dx^\mu,Y,P,\mph,c^{+})\,, \qquad \Omega_0\Phi^{red}=0\,, \qquad \Phi^\red\sim \Phi^\red+(c^+)^2\chi\,. 
\end{gathered}
\end{equation}
In terms of component fields this reduction amounts to the elimination of $f_{++}$ through its equations of motion $f_{++}=\qcommut{P\cdot\mph}{f_+}$. Upon the reduction the system takes the form
\besubeqs
\label{linear-par-red}
\begin{align}
    D_0a&=0\,, &\delta a&=D_0 \xi\,,\\
    D_0 f_+&=\qcommut{a}{F_{+}^0}\,, &\delta f_+&=\qcommut{F_{+}^0}{\xi}\,,\\
    \qcommut{F_-^0}{f_+}&=\qcommut{F_{--}^0}{f_+}=\qcommut{F_{0}^0}{f_+}-f_+=0\,, & 
    \qcommut{F^0_-}{a}&=\qcommut{F_{--}^0}{a}=\qcommut{F_{0}^0}{a}=0\,,
    \end{align}
\esubeqs
where the gauge parameter $\xi$ also satisfies $\qcommut{F_-^0}{\xi}=\qcommut{F_{--}^0}{\xi}=\qcommut{F_{0}^0}{\xi}=0$. Note that the equations involving $F_{--}$ are consequences of those with $F_-$ and hence can be dropped. In its turn the above linear system can be arrived at by reformulating~\eqref{linearized-hook} in the parent form so that it is equivalent to the off-shell version of the free Type-B theory. 

The above homological arguments showing that the linearized parent system is equivalent to the multiplet of linear fields is a straightforward generalization of those given in~\cite{Grigoriev:2012xg}, which in turn are substantially based on~\cite{Barnich:2006pc,Grigoriev:2006tt} (see also~\cite{Bekaert:2009fg,Alkalaev:2009vm}).

%%%%%%%%%%%%%%%%%%%%%%%%%%%%%%%%%%%%%%%%%%%%%%%%%%%%%%%%%%%%%
\subsection{On-shell System}
\label{sec:onshell}
%%%%%%%%%%%%%%%%%%%%%%%%%%%%%%%%%%%%%%%%%%%%%%%%%%%%%%%%%%%%%
Now using off-shell system \eqref{Bsystem} as a starting point we take into account the gauge symmetry \eqref{ext2} related to the redefinition of the constraints and let structure functions \eqref{ext1} to vary. More precisely, if we view equations \eqref{Bsystem} as consistency conditions 
for the constrained system with constraints $\d_\mu+A_\mu$, $F_i$ and hence identify $dx^\mu$ as ghost variables, one would allow redefinitions of the constraints $A_\mu$ as well. Nevertheless, it is convenient not to redefine  $F_i$ through $\d_\mu+A_\mu$ because on one hand this spoils geometric interpretation and on the other hand such redefinitions are not really needed
because any redefinition of the $F_i$ constraints in the ambient space can be lifted to a system of the form \eqref{Bsystem}.

In this way we arrive at the system: 
\besubeqs
\label{full-parent}
\begin{align}
 dA-\half\qcommut{A}{A}&= u^i\star F_i\,, &
 \delta A&=d\xi-\qcommut{A}{\xi}+\lambda^j\star F_j \,,\\
 dF_i-[A,F_i]_\star&= u^j_i\star F_j\,, &\delta F_i&=\qcommut{\xi}{F_i}+ \lambda^j_i \star F_j\,,\\ \qcommut{F_i}{F_j}-C_{ij}^kF_k&= u_{ij}^k \star F_k\,.
\end{align}
\esubeqs
Here, the new $u$-fields account for the variation of the structure constants and are not considered dynamical. So the system is understood as equations on $F_i$, $A_\mu$ which say that $F_i,A_\mu$ are such that equations can be satisfied with some $u$. A general systematic procedure to arrive at the system~\eqref{full-parent} is to employ the BRST formalism and AKSZ sigma models, see~\cite{Grigoriev:2006tt,Grigoriev:2012xg} for more detail. See also~\cite{Alkalaev:2014qpa} for more detail on the  algebraic interpretation of equations~\eqref{full-parent}. In so doing the $u$-fields appear at the equal footing with other fields. Note that~\eqref{vacuum-parent} is a particular solution of the above on-shell system.

In what follows we refer to system~\eqref{full-parent} defined on $(d+1)$-dimensional space as to {\it on-shell system}. Unless otherwise specified  it is assumed that the theory is understood around the vacuum where $F=F^0_i$, with $F^0_i$ defined in~\eqref{vacuum-parent}. This on-shell system is our proposal for the Type-B theory.

\paragraph{Linearization.}
The linearized equations are obtained from \eqref{full-parent} as before, i.e. by replacing $A\rightarrow A^0+a$ and $F_i\rightarrow F_i^0+f_i$ and picking the terms linear in $a$ and $f_i$:%
\besubeqs
\label{linearized-pA}
\begin{align}
    D_0a&=u^i \star  F^0_i\,, & \delta a&=D_0\xi+ \lambda^i \star F^0_i\,,\\
    D_0f_i-[a,F_i^0]_\star&= u^j_i \star F^0_j\,, & \delta f_i&= [\xi,F_i^0]_\star+ \lambda^j_i \star F^0_j\,, \\
    [F^0_i,f_j]_{\star ,\pm}-(ij)-C_{ij}^k f_k&=u^k_{ij}\star F^0_k\,,
\end{align}
\esubeqs%
where $A^0$, $F^0_i$ is a vacuum solution, e.g. \eqref{vacuum-parent}, and $D_0\bullet \equiv d\bullet -[A^0,\bullet]$ is the background covariant derivative.

Now we are going to make use of the $\lambda$ gauge symmetry in order to explicitly formulate \eqref{linearized-pA} as an on-shell system. Suppose that we have succeeded to identify such a class of functions in $Y,P,\mph$ that, by using $\lambda$-gauge transformations, fields $a,f_i$ can be made totally traceless, i.e. belonging to the kernel of the following operators:
\begin{equation}
\label{trace-operators}
    \d_Y\cdot\d_Y\,, \qquad \d_Y\cdot\d_P\,, \qquad\d_P\cdot\d_P\,, \qquad \d_Y\cdot\d_\mph\,, \qquad \d_P\cdot\d_\mph\,.
\end{equation}
Then the right hand sides in~\eqref{linearized-pA} vanish and the system takes the form \eqref{linearized-p}. Note that the above condition is a parent formulation counterpart of the equations~\eqref{linearized-hook-eoms+traces}.

As before linearized equations of motion and gauge symmetries can be encoded in~\eqref{BRST-form}, where now $\Phi,\Xi$ belong to the kernel of~\eqref{trace-operators}. All the steps leading to the system~\eqref{linear-par-red} remain unchanged except that now one needs to employ the known results~\cite{DeBie:2007zoh,Luo-osp:2010,Lavicka:2017} on the structure of polynomial algebras in the supersymmetric case in order to make sure that the $\Delta_{-1}$ cohomology is concentrated in degree zero in $c^{--},c^0,c^-$-ghosts. In this way one arrives at the system~\eqref{linear-par-red} but with $a,f_+$ belonging to the kernel of \eqref{trace-operators}. This system can be shown to be a parent reformulation
of the ambient space system~\eqref{linearized-hook}, \eqref{linearized-hook-eoms+traces}. To summarize, under the assumptions made we have shown that the on-shell system \eqref{full-parent} reproduces the free Type-B theory introduced in Section~\bref{sec:ads-free} upon linearization around the vacuum solution~\eqref{vacuum-parent} describing AdS space.

Note that the linear parent formulation in terms of BRST operator~\eqref{BRST-form}, \eqref{BRST-op} with $\Phi,\Xi$ in the kernel of \eqref{trace-operators} can be used to arrive directly to the metric-like approach of section \bref{sec:ads-free} by performing a suitable homological reduction, see e.g.~\cite{Barnich:2006pc} for the analogous procedure in the case of totally symmetric fields. Homological methods may also be used to formulate the theory in terms of physical degrees of freedom only, see e.g.~\cite{Barnich:2005ga,Alkalaev:2008gi} for similar analysis of higher spin fields in Minkowski space.

%%%%%%%%%%%%%%%%%%%%%%%%%%%%%%%%%%%%%%%%%%%%%%%%%%%%%%%%%%%
\subsection{Functional Class} 
%%%%%%%%%%%%%%%%%%%%%%%%%%%%%%%%%%%%%%%%%%%%%%%%%%%%%%%%%%%
As it was already noted, an interpretation of the theory of background fields crucially depends on the choice of vacuum as physics over different vacua can be very much different. For example, it makes sense to expand $\mathrm{tr}\log F$ over $i(\gamma\cdot \pl_x)$ (and it is also known how to do that), but it maybe hard to make any sense out of a randomly picked background $F$. The situation gets worse when going from boundary to bulk as higher spin theories are hard to interpret within the usual field theory framework. Depending on a functional class chosen for $A$ and $F_i$ one can end up with an empty system (all solutions are pure gauge). As we just seen in the previous section, in order to reproduce linearized theory one needs to assume that dependence of fields on the auxiliary variables $Y,P,\mph$ is such that some of the equations admit unique solutions and, at the same time, certain gauge conditions are reachable. Therefore, a proper choice of functional class is a crucial ingredient of the setup.

The definition of the functional class is almost identical to the one in \cite{Bekaert:2017bpy}. The functional class  $\mathfrak{C}$ is that of polynomials in $P$ and $\Mph$ with coefficients that are formal power series in $Y$. Also we need  $\mathfrak{C}$ to be of finite trace order, i.e. for any $f(Y,P,\mph)\in \mathfrak{C}$ there exists an $\ell\in\mathbb N$ such that 
\begin{align}
{(\d_Y\cdot \d_Y)^\ell f=0\,.}\label{ftr}
\end{align}
By definition, any given function in $\mathfrak{C}$ can be decomposed as
\begin{align}
\label{decom-tr}
f=f_0+f^i_{1}\star F^0_i + f^{ij}_{2} \star F^0_i\star F^0_j+\ldots\,, \qquad\qquad f_{n}\text{ -- totally traceless}
\,,
\end{align}
such that the number of terms is finite. By totally traceless we mean that all possible traces in $Y$, $P$, $\Mph$ vanish, i.e. $f_n$ is in the kernel of the operators~\eqref{trace-operators}. It follows that if $a,f_i,\xi,\lambda$ in~\eqref{linearized-pA} belong to the functional class (and hence admit decomposition~\eqref{decom-tr}), then 
the $\lambda$ gauge symmetry can be used to set $a,f_i$ totally traceless and hence the assumption made in section~\bref{sec:onshell} is satisfied. One concludes that with this functional class the linearized on-shell system~\eqref{linearized-pA} is well defined and indeed describes linearized Type-B theory.

For further convenience let us define a projector onto the traceless part: 
$
{\Pi f=f_0}
$. 
It is also important that $\mathfrak{C}$ is a module over polynomials in $Y,P$. Hereafter we assume that $A^0$ as well as fluctuations $a,f_i$ belong to $\mathfrak{C}$. Let us note that $\mathfrak{C}$ is apparently not closed under multiplication and does not yet allow us to immediately discuss higher orders, which is not independent of the locality problem discussed in section \bref{sec:conclusions}.

%%%%%%%%%%%%%%%%%%%%%%%%%%%%%%%%%%%%%%%%%%%%%%%%%%%%%%%%%%%
\subsection{Higher Spin Flat Backgrounds}
\label{sec:hsflat}
%%%%%%%%%%%%%%%%%%%%%%%%%%%%%%%%%%%%%%%%%%%%%%%%%%%%%%%%%%%
First, we define the Type-B higher spin algebra $\hsb$ as the centralizer of $F^0_i$ modulo $F^0_i$:
\begin{equation}
    \hsb=\{\hsel \in \mathfrak{C}:\qcommut{\hsel }{F^0_i}=0\,, \quad  \hsel \sim \hsel +\lambda^i\star F_i^0\}\,,
\end{equation}
where $F^0_i$ are given by~\eqref{vacuum-parent}. In particular, elements of $\hsb$ are polynomials in $Y$. Indeed, $\qcommut{\hsel }{F^0_{++}}=(P\cdot\d_Y) \hsel=0$ together with the assumption that all elements are polynomial in $P$ imply that $\hsel $ is a polynomial in $Y$ as well. Furthermore, we can assume representatives $\hsel$ totally traceless, i.e. $\Pi(\hsel )=\hsel$, using the equivalence relation.
The above higher spin algebra is, by definition, the same as the symmetry algebra of the Dirac equation \cite{Nikitin1991fer}, i.e. the symmetry of the $i(\gamma\cdot p)$ vacuum of section \bref{sec:fermion-mink}. Almost identical oscillator realization as arises here was given in \cite{Vasiliev:2004cm} (the difference is the absence of $V^A$-shift in the vacuum $F^0_i$). 

The relevance of $\hsb$ can be seen by considering more general vacuum solutions. Namely we take $F^0_i$ as in~\eqref{vacuum-parent}
but do not restrict $A^0$ to be at most quadratic in $P,Y,\Mph$. Assuming that $A_0$ belongs to $\mathfrak{C}$ it can be represented
as $A^0_0+A^{0i}_1\star F^0_i+...$. Equations of motion \eqref{full-parent} of the on-shell system then imply
\begin{align}
\label{hsflat}
\qcommut{A^0_0}{F^0_i}&=0\,, && dA^0_0=\Pi ( A^0_0\star A^0_0)\,.
\end{align}
Note that $\Pi$ in the second equation is well-defined thanks to $A^0_0$ being polynomial in $Y$, which in turn is a consequence of the first equation. We conclude that $A^0_0=\Pi(A^0)$ is a flat connection of Type-B higher spin algebra $\hsb$. 

Consequently, we have arrived at a family of solutions to the on-shell system~\eqref{full-parent}
which correspond to flat connections of the Type-B higher spin algebra. These backgrounds are maximally symmetric and the symmetries are in one-to-one with the algebra $\hsb$ and can be identified with (mixed-symmetry) Killing tensors. Indeed, the gauge symmetries preserving the vacuum solution are determined by 
\begin{align}
    d\epsilon^0&=[A^0,\epsilon^0]_\star+ \lambda^i\star F^0_i\,, & [F^0_i,\epsilon^0]_\star+\lambda^j_i\star F^0_j &=0\,.
\end{align}
We can again decompose $\epsilon^0$ into the trace part that is proportional to $ F^0_i$ and the traceless part. The trace part fixes the $\lambda$'s up to an equivalence, while the traceless part is covariantly constant with respect to $A^0$. Therefore, the global symmetry algebra is the Type-B higher spin algebra $\hsb$.

Let us now study the linearized system~\eqref{linearized-pA} taking as a background solution $F^0_i$ from \eqref{vacuum-parent} and taking $A^0$ to be a flat connection of $\hsb$. Assuming that fluctuations belong to $\mathfrak{C}$ we can exploit $\lambda$-symmetry to set  $\Pi(a)=a$ and $\Pi(f_i)=f_i$ and explicitly apply the trace projector $\Pi$
to all the equations. The resulting system is identical to~\eqref{linearized-p} except that all fields are traceless and $\Pi$ explicitly enters the expression for the covariant derivative $D_0$.

It turns out that all the steps of the analysis performed in Section~\bref{sec:off-shell} do not depend on the particular choice of $A^0$ as all the operators involved in the reduction are determined by $F^0_i$. Upon elimination of all the fields save for $a,f_+$ the equations of motion and gauge symmetries take the form
\besubeqs\label{linonshell}
\begin{align}
    D_0a&=0\,,  &D_0f_+&= \qcommut{a}{F_{+}^0}\,, \\
    \delta a &=D_0\xi\,,&\delta f_+&=\qcommut{\xi}{F_{+}^0}\,,\\
    \qcommut{F^0_-}{f_+}=\qcommut{F^0_-}{a}&=\qcommut{F^0_-}{\xi}=0\,, &
   \qcommut{F^0_{0}}{f_+}-f_+&=\qcommut{F^0_{0}}{a}=\qcommut{F^0_{0}}{\xi}=0\,,
\end{align}
\esubeqs
where $D_0\bullet\equiv d\bullet-\Pi[A^0,\bullet]_\star $. 
This gives a concise formulation of the multiplet of hook-type fields propagating on the background of generic flat connection of the respective higher spin algebra.

%%%%%%%%%%%%%%%%%%%%%%%%%%%%%%%%%%%%%%%%%%%%%%%%%%%%%%%%%%%%%
\subsection{Boundary Values and Holographic Reconstruction}
%%%%%%%%%%%%%%%%%%%%%%%%%%%%%%%%%%%%%%%%%%%%%%%%%%%%%%%%%%%%%

The on-shell system~\eqref{full-parent} was constructed starting from the ambient formulation of the background fields for a conformal fermion in $d$-dimensions. The important step was to reinterpret the ambient system in a different way by considering it in the vicinity of the hyperboloid rather than hypercone. In the parent formulation~\eqref{full-parent} this is made manifest by explicitly defining the theory on $AdS_{d+1}$ space and specifying the natural vacuum in terms of the AdS-like compensator field $V$, $V^2=-1$ together with a compatible $o(d,2)$-connection.

If the AdS system is formulated in the ambient space approach the boundary values of the AdS fields are described by the same system considered in the vicinity of the hypercone rather than the hyperboloid. In the parent formulation, this simply corresponds to considering the same system to be defined on the $d$-dimensional conformal space  and taking the conformal version of the connection and the compensator field entering the vacuum solution. More specifically, in the vacuum solution~\eqref{vacuum-parent} one should take  $\omega,V$ such that $V^2=0$, $\omega$ is a flat $o(d,2)$-connection and  $\nabla V^B$ has maximal rank (which is $d$). In this way one indeed recovers correct boundary values  as was explicitly verified for free totally symmetric fields in~\cite{Bekaert:2012vt,Bekaert:2013zya}, and for mixed-symmetry fields in~\cite{Chekmenev:2015kzf}.\footnote{{Mention also somewhat related unfolded approach to boundary values proposed in~\cite{Vasiliev:2012vf}, where the boundary values are described by the free differential algebra associated to the bulk fields. However, the approach of~\cite{Vasiliev:2012vf} does not employ the ambient space construction and compensator field  which are the crucial ingredients of the present construction.}}

Applying this procedure to the on-shell system~\eqref{full-parent} we arrive at the parent reformulation of the theory of background fields for the Dirac fermion in $d$-dimensions. This in turn can be explicitly shown to reproduce the system \eqref{constraints} that describes background fields for the Dirac fermion in $d$-dimensions upon linearization. 
In this way we confirm that the on-shell system~\eqref{full-parent} indeed has the correct boundary values.

What is less trivial is to observe that the passage between the bulk and the boundary can be explicitly done for the system linearized over generic higher spin flat connection. To see this suppose we are given a higher spin flat connection $A^0$, which is defined in the bulk. Because $A_0$ is a polynomial in $Y,P,\Mph$
it can be rewritten in terms of new variables $Y^\prime=V+Y,P,\Mph$.  Moreover, the algebraic constraints $\qcommut{A^0}{F^0_i}=0$ on $A_0$ can be also rewritten in terms of $Y^\prime,P,\Mph$ and hence the compensator field $V$ does not explicitly enter. After this $A_0$ can be pulled back to the boundary, giving a higher spin flat connections there. Then changing variables to $Y=Y^\prime-V_{conf},P,\Mph$ but now with $V_{conf}$ satisfying $V_{conf}^2=0$ one arrives at the vacuum solution in the boundary theory corresponding to a flat connection. In this way we can relate the bulk fields defined over a higher spin flat background and their associated boundary values.

It is clear from the above discussion that the on-shell system~\eqref{full-parent} is in some sense a consistent lift of the theory of background fields for the Dirac fermion in $d$-dimension to the $(d+1)$-dimensional bulk. By this lift  the off-shell background fields defined on the boundary become "on-shell" in the bulk.  Note that even the very possibility of such a lift (in contrast to the reverse procedure of extracting the theory of boundary values from a given bulk theory) heavily relies on the proper ambient space formulation of the original theory in terms of the underlying constrained system.
Additional important, though somewhat technical step, is the parent reformulation which makes it possible to fine-tune functional classes and to give the bulk theory interpretation in terms of fields explicitly defined in the bulk.

%%%%%%%%%%%%%%%%%%%%%%%%%%%%%%%%%%%%%%%%%%%%%%%%%%%%%%%%%%%%%
\section{Unfolded Form and Hochschild Cocycle}
\label{sec:fda}
%%%%%%%%%%%%%%%%%%%%%%%%%%%%%%%%%%%%%%%%%%%%%%%%%%%%%%%%%%%%%

We have already shown that linearization of the Type-B equations reproduce the expected free field content. This is an important check, although it does not probe the structure of interactions. Even a linearized system encodes nonlinear structures if it is known for a sufficiently general background. It turns out that free equations over higher spin flat backgrounds, which are maximally symmetric, contain all the nontrivial information in some sense. This can be made manifest by reformulating the system as a minimal Free Differential Algebra (FDA).

On shell system \eqref{full-parent} already has a form of FDA with constraints or, which is almost the same, AKSZ sigma model and is entirely determined by the underlying Q-manifold. This is most easily seen for the off-shell version of the theory and equally well applies to the on-shell, at least at the linearized level. When the system is reformulated in terms of metric-like fields of section \bref{sec:ads-free} such structures become hidden but can be recovered~\cite{Barnich:2010sw} from the BRST formulation of the system.

Searching for higher spin gravity equations in the form of FDA underlies the unfolded approach~\cite{Vasiliev:1988sa}. In contrast to~\eqref{full-parent} the unfolded formulation typically operates with the minimal FDA without constraints (though the Vasiliev system~\cite{Vasiliev:1990cm,Vasiliev:2003ev} nevertheless involves certain non-minimal extension needed to package the series in curvatures into concise expressions). Once the theory is linearized over a higher spin flat background, one can systematically eliminate generalized auxiliary fields (contractible pairs for the underlying $Q$-manifold) and extract the minimal FDA, which is naturally formulated in terms of the higher spin algebra (symmetry of the vacuum) and its Hochschild cocycle. These two completely fix the minimal FDA \cite{Sharapov:2017yde,Sharapov:2017lxr} and also allow one (at least in principle) to reconstruct the entire nonlinear system, as a formal perturbative expansion in curvatures.

%%%%%%%%%%%%%%%%%%%%%%%%%%%%%%%%%%%%%%%%%%%%%%%%%%%%%%%%%%%%%
\subsection{FDA over AdS Background}
\label{sec:freefda}
%%%%%%%%%%%%%%%%%%%%%%%%%%%%%%%%%%%%%%%%%%%%%%%%%%%%%%%%%%%%%
For a generic linear system there is a systematic procedure to derive minimal unfolded formulation from the parent one~\cite{Barnich:2004cr,Barnich:2010sw}. In BRST terms this amounts to reducing the theory to the cohomology of the target space BRST operator. For the system~\eqref{linonshell}, this is the term $c^+\qcommut{F^0_+}{\bullet}$ in the BRST operator~\eqref{omega-red}. The reduction is then algebraically identical to the analogous reduction for totally symmetric fields, see~\cite{Barnich:2004cr,Barnich:2006pc} for more detail.

Instead of performing the reduction in homological terms we do it explicitly in terms of component fields, and first for the AdS vacuum.\footnote{{In the case of totally-symmetric fields in flat space the component form of the analogous reduction was discussed in~\cite{Vasiliev:2005zu}.}} By sightly changing notation,~\eqref{linonshell} can be rewritten as ($ad_-\equiv [F_-^0,\bullet]$, $ad_+\equiv [F_+^0,\bullet]$, $ad_0\equiv [F_0^0,\bullet]$)
\begin{align}
    D_0 a&=0\,, & \delta a&=D_0\xi\,,\\
    D_0 f_+ &=-ad_+ a\,,  &\delta f_+&=-ad_+ \xi\,, 
\end{align}
where the fields and gauge parameters belong to 
\begin{align}\label{othercond}
    ad_{-}a&=ad_{-}\xi=ad_{-}f_+=0\,, & (ad_{0})a&=(ad_{0})\xi=(ad_{0}-1)f_+=0\,.
\end{align}
It is clear that a part of the gauge symmetry is still algebraic and allows us to further gauge away certain components of $f_+$.

It is useful to decompose the space of "functions" in $P,Y,\Mph$ into $\mathrm{Ker}(ad_+)\oplus \mathrm{Coim}(ad_+)$. An alternative decomposition that we also employ is $\mathrm{Im}(ad_+)\oplus \mathrm{Coker}(ad_+)$. Note that both decompositions are not canonical. Let us then split the fields 
and gauge parameters accordingly as:
\begin{align}\label{gaugeunfld}
\begin{aligned}
    \xi&= \epsilon +\tilde{\xi}\,,\qquad\qquad&& \epsilon\in \mathrm{Ker}(ad_+)\,,\qquad \tilde{\xi}\in\mathrm{Coim}(ad_+)\,,\\
    a&= \omega+\tilde{a}\,,\qquad\qquad&& \omega\in \mathrm{Ker}(ad_+)\,,\qquad \tilde{a}\in\mathrm{Coim}(ad_+)\,,\\
    f_+&= \tilde{f}+C\,,\qquad\qquad && \tilde{f}\in \mathrm{Im}(ad_+)\,,\qquad C\in \mathrm{Coker}(ad_+)\,.
\end{aligned}
\end{align} 
It follows that we can gauge away $\tilde{f}$ with the help of $\tilde{\xi}$. Then, the system can be written as
\begin{align}
    D_0 \omega&=-D_0 \tilde{a}\Big|_{\mathrm{Ker}}\,, &
    D_0 C\Big|_{\mathrm{Im}} &=ad_+ \tilde{a}\,,&
    D_0 C\Big|_{\mathrm{Coker}}&=0\,,
\end{align}
where we projected the equation for $C$ onto the two subspaces since $D_0$ does not preserve our (non-canonical) gauge choice. In the last step we can express $\tilde{a}=ad_+^{-1}(...)$ from the second equation (since $ad_+ \tilde{a}$ can be inverted on the $\mathrm{Im}(ad_+)$) and substitute the solution into the first equation:
\begin{align}
    D_0\omega&=-D_0\, ad_+^{-1}D_0 C\Big|_{\mathrm{Ker}}\,,&
    D_0 C\Big|_{\mathrm{Coker}}&=0\,.
\end{align}

The first equation can be simplified if we assume that the decomposition \eqref{gaugeunfld} preserves Lorentz covariance. Provided such a choice has been made the Lorentz-covariant derivative $\nabla$ part of $D_0$ drops out and only the translation generators $P^m$ contribute.\footnote{The translation generators are defined as $P^m=T^{m,{-1}}$, i.e. as the components along $V^A$, $T^{AB}V_B$. Likewise, $L^{mn}=T^{mn}$ and we rewrite $D_0\bullet=\nabla\bullet- h^m [P_m,\bullet]$, where $h^m$ is the $AdS_{d+1}$ vielbein.} Therefore, the full system can be rewritten as
\begin{align}
    D_0 \omega&=-h_m\wedge h_n\, ad(P^m)\, ad_+^{-1}ad(P^n)\, C\Big|_{\mathrm{Ker}}\,, &
    D_0 C\Big|_{\mathrm{Coker}}&=0\,,\label{freecocycle}
\end{align}
where $ad(P^m)\equiv[P^m, \bullet]$. This is the right structure of free equations describing mixed-symmetry fields \cite{Alkalaev:2003qv,Barnich:2006pc,Boulanger:2008up,Boulanger:2008kw,Skvortsov:2009nv,Alkalaev:2009vm,Skvortsov:2009zu,Alkalaev:2011zv}.

The above arguments do not show that the r.h.s. of $D_0\omega=$ in \eqref{freecocycle} is nontrivial, but we have already reproduced the correct free equations in section \bref{sec:onshell}. If the r.h.s. of $D_0\omega=$ were trivial,  $\omega$ would be equivalent to a linearized flat connection, i.e. pure gauge. Here we should note that in the unfolded formulation just constructed the free fields of section \bref{sec:ads-free} reside as particular components of $\omega$ and hence getting a non-trivial equation for $\omega$ is crucial.

Finally, we note that, despite the appearance, $\omega$ and $C$, are isomorphic linear spaces: both $\omega$ and $C$ can be decomposed into irreducible Lorentz tensors in the fiber space to find a doubled set of tensors of type (i.e. the independent Taylor coefficients in $y^m$, $p^n$, $\mph^k$)
\begin{align}\label{spectrum}
    \bigoplus_{r,t,p}\parbox{60pt}{{\bep(60,60)\unitlength=0.38mm%
    \put(0,40){\RectARowUp{4}{$t$}}%
    \put(0,50){\RectARowUp{7}{$r$}}%
    \put(0,40){\YoungA}\put(0,50){\YoungA}%
    \put(40,40){\YoungA}\put(60,50){\YoungA}%
    \put(0,0){\YoungA}%
    \put(0,0){\RectT{1}{4}{\TextCenter{$p$}}}\eep}}
\end{align}
This isomorphism will be important below.

%%%%%%%%%%%%%%%%%%%%%%%%%%%%%%%%%%%%%%%%%%%%%%%%%%%%%%%%%%%%%
\subsection{FDA over Higher Spin Flat Background}
%%%%%%%%%%%%%%%%%%%%%%%%%%%%%%%%%%%%%%%%%%%%%%%%%%%%%%%%%%%%%
As we showed in section \bref{sec:hsflat}, the Type-B system \eqref{full-parent} admits more solutions than just $AdS_{d+1}$.  Without any change in the logic and in the formulas we can linearize the system over any flat connection:
\begin{align}
    dA^0&=A^0\ast A^0\,, && A^0\in \hsb\,,\label{flatagain}
\end{align}
where, by somewhat abusing notation, we work in terms of fields taking values in the higher spin algebra $\hsb$ understood as an abstract associative algebra with the product denoted by $\ast$. This is in contrast to \eqref{hsflat}, where $\hsb$ was realized as a certain subquotient of the Clifford-Weyl algebra. Here we would like to stress that it makes sense to work with $\hsb$ as an abstract associative algebra.\footnote{The realization via the Clifford-Weyl algebra given in section \bref{sec:hsflat} is not the only one. One can consider symmetries of the Dirac operator \cite{Nikitin1991fer}, quasi-conformal realization \cite{Fernando:2015tiu}, enveloping algebra realization \cite{Boulanger:2011se} etc. All of them give the same $\hsb$.}

The steps from the previous section can literally be repeated. Indeed, the split \eqref{gaugeunfld} appeals to $ad_+$, i.e. $F_+^0$, and not to a particular choice of $A^0$. Recall also that $a$, $f_+$ and $\xi$ are assumed to be traceless, e.g. $\Pi(a)=a$. The only change is to replace the AdS background derivative $D_0$ with the $A^0$ background derivative $D_0\bullet\equiv d\bullet-\Pi[A^0,\bullet]_\star$, where the $\Pi$-projector needs to be added. We end up with\footnote{Recall that the split \eqref{gaugeunfld} commutes with $d$ and for that reason $d$ drops out.}
\begin{align}
    D_0 \omega&=-\, ad(A^0)\, ad_+^{-1}ad(A^0)\, C\Big|_{\mathrm{Ker}}\,, \label{freecocycleA} &
     dC&=ad(A^0) C\Big|_{\mathrm{Coker}}\,, 
\end{align}
where $ad(A^0)\bullet \equiv \Pi[A^0,\bullet]_\star$.

Recall that $\omega$ belongs to the higher spin algebra $\hsb$ due to \eqref{othercond} and \eqref{gaugeunfld}. It is less obvious that $C$ can also be thought of as an element of $\hsb$. To be precise, we claim that there exists a map $\rho$ from the space where $C$ takes values, \eqref{othercond}, \eqref{gaugeunfld}, to $\hsb$ such that for all $\hsel \subset \hsb$
    $$\begin{tikzcd}[row sep=large, column sep=huge]
    C\ar[d, ""]\ar[r, "\rho"] & \Cone=\rho(C)
    \in \hsb \ar[d, ""]\\
    \Pi[\hsel,C]_\star \ar[r, "\rho"]& \hsel\ast\Cone-\Cone\ast\pi(\hsel)
    \end{tikzcd}$$  
Here, $\pi$ in \eqref{Ceq} is an automorphism of the higher spin algebra induced by the automorphism of the anti-de Sitter subalgebra $\pi(L_{mn})=L_{mn}$ and $\pi(P_m)=-P_m$. This automorphism is required to reproduce the right structure of the free equations,\footnote{This can \maxim{be} easily proved using the techniques from \cite{Iazeolla:2008ix,Basile:2018dzi}, we avoid giving the technicalities here.} which we obtained in sections \bref{sec:typeb} and \bref{sec:freefda}. 
One argument in favour of $\rho$ is that $\omega$ and $C$ are isomorphic as Lorentz algebra modules, \eqref{spectrum}. Moreover, such $\rho$ clearly exists for $\hsel\in o(d,2)\subset \hsb$.

Now we are ready to reformulate everything in the language of higher spin algebras. Eqs. \eqref{flatagain}, \eqref{freecocycleA} for $A^0$, $\omega$ and $C$ can be recognized to have the following general structure
\besubeqs\label{linfluct}
\begin{align}
dA^0&=A^0\ast A^0\,,\\
d\omega&=A^0 \ast \omega +\omega\ast  A^0+\mathcal{V}(A^0,A^0,\Cone)\,,\label{omegaeq}\\
d\Cone&=A^0 \ast \Cone-\Cone\ast \pi( A^0)\label{Ceq}\,,
\end{align}
\esubeqs
where by abusing the notation again we assumed that $A^0$, $\omega$ take values in $\hsb$. Here the vertex $\mathcal{V}$ is some trilinear map that appears in the first of \eqref{freecocycleA}. The term $A^0 \ast \omega +\omega\ast A^0$ comes from $D_0\omega$. 

Equations \eqref{linonshell} of section \bref{sec:typeb} or \eqref{flatagain}, \eqref{freecocycleA}, \eqref{linfluct} describe propagation of higher spin fields of the Type-B theory over an arbitrary higher spin flat background $A^0$. Indeed, this is how we obtained them from the Type-B theory as a linear approximation. As we already discussed this result is much stronger than just the check that the free Type-B over $AdS_{d+1}$ is reproduced upon linearization: knowing how fields propagate on sufficiently general backgrounds should contain some information about interactions thereof too. 

The system that \eqref{linfluct} is a linearization of should have the following schematic form
\besubeqs
\label{linearC}
\begin{align}
&d\omega=\omega\ast\omega+\mathcal{V}(\omega,\omega,\Cone)+\mathcal{O}(\Cone^2)\,,\\
&d\Cone=\omega\ast \Cone-\Cone\ast \pi(\omega)+\mathcal{O}(\Cone^2)\,.
\end{align}
\esubeqs
Consistency of \eqref{linfluct} or consistency of \eqref{linearC} to order $\Cone$ imposes some restrictions on $\mathcal{V}$. To identify the restrictions and to find the structure of $\mathcal{V}$ it is convenient to consider a slightly more general setting, where fields $\omega,\Cone$ take values in a matrix algebra, e.g. $u(M)$ (the appropriate reality condition has to be assumed). This step is a straightforward generalization of the analogous step known in the context of Vasiliev approach~\cite{Vasiliev:1988sa}. At the level of parent formulation~\eqref{full-parent} this corresponds to taking
all the fields $A$ and $F_i$ also matrix-valued which does not spoil consistency of the system. In the CFT dual description, this would correspond to taking $U(M)\times U(N)$ fermions and imposing the $U(N)$ singlet constraint, and hence having global $U(M)$ symmetries left, so that $u(M)\times \hsb$ gets gauged in the bulk. 

Given such matrix extensions  the consistency of the system implies (see \cite{Sharapov:2017yde} for more detailed discussion) that $\mathcal{V}$ have the following structure:
\begin{align}\label{firstorder}
\mathcal{V}(\omega,\omega,\Cone)=\Phi(\omega,\omega)\ast \pi(\Cone)\,,
\end{align}
where $\Phi(\bullet,\bullet)$ is a Hochschild two-cocycle of the higher spin algebra:
\begin{align}\label{sptwoHochschildA}
    \hsel\ast \Phi(\hselb,\hselc)+\Phi(\hsel\ast \hselb,\hselc)-\Phi(\hsel,\hselb\ast \hselc) +\Phi(\hsel,\hselb)\ast \pi(\hselc)=0\,, \qquad \hsel,\hselb,\hselc\in \hsb\,.
\end{align}
Heuristically, $\mathcal{V}$ is needed to avoid triviality of flat connections. The deviation from flatness is controlled by $\Cone$ and higher orders in $\Cone$ may be required to make the full system \eqref{linearC} formally consistent. All these structures are closely related to deformation quantization and formality \cite{Sharapov:2017yde,Sharapov:2017lxr,Sharapov:2018hnl}, which is another reason for the qualifier 'formal' in the title.

It follows the Type-B equations we propose do reproduce a nontrivial Hochschild two-cocycle. It is important that the linearized system \eqref{linonshell} and its reductions \eqref{flatagain}, \eqref{freecocycleA} and rewriting \eqref{linfluct} still capture the full structure of $\mathcal{V}(\omega,\omega,\Cone)$, i.e. all three arguments can be arbitrary elements of the higher spin algebra (when $A^0$ takes values in just $o(d,2)\subset \hsb$ a significant part of $\mathcal{V}$ is lost). 

Finally, under quite general assumptions it can be shown \cite{Sharapov:2017yde,Sharapov:2017lxr} that the deformation induced by $\Phi(\bullet,\bullet)$ is unobstructed.\footnote{An interesting issue that we ignore is that different types of fermions on the boundary can lead to slightly different deformations/spectrum of fields, see e.g.  \cite{Giombi:2016pvg,Gunaydin:2016amv,Basile:2018dzi}. More detail will be given elsewhere. } 
More precisely, it can be shown that the higher order terms making \eqref{linearC} consistent to all orders exist and are unique up to the natural equivalence, so that the formally consistent unfolded system~\eqref{linearC} can be constructed to all orders (at least in principle) starting from the on-shell system~\eqref{full-parent} proposed in this work. This supports our claim that the on-shell system ~\eqref{full-parent} describes the Type-B higher spin gravity, at least at the formal level.

%%%%%%%%%%%%%%%%%%%%%%%%%%%%%%%%%%%%%%%%%%%%%%%%%%%%%%%%%%%%%
\section{Conclusions and Discussion}
\label{sec:conclusions}
%%%%%%%%%%%%%%%%%%%%%%%%%%%%%%%%%%%%%%%%%%%%%%%%%%%%%%%%%%%%%

From AdS/CFT correspondence perspective the problem of higher spin gravity can be understood as a problem of constructing a dual of a given free CFT restricted to various bilinear operators to be understood as single-trace ones. In the paper we proposed the following general recipe to construct higher spin gravities: (i) instead of dealing with the generating functional of correlators of single-trace operators one can study a more general question of how to couple them to an arbitrary higher spin background $\tildef$; (ii) the power of the higher spin symmetry is that the generating functional $W[\tildef]$ turns out to be completely fixed by the non-abelian infinite-dimensional gauge symmetries $\delta \tildef=\pl \xi+...$; (iii) the theory of background fields, i.e. field content and non-abelian gauge symmetries $\delta \tildef=...$, can be understood as  quantum constrained system that gives a first-quantized description of a given free CFT; (iv) the constrained system can be reformulated in the ambient space where it gives the background fields on the projective cone $X^2=0$; (v) the same system can be then considered on the hyperboloid $X^2=-1$ where it describes non-linear equations whose free field approximation gives the same sources $\tildef$ as boundary values. 

Although most of the above steps are not entirely new and have been already applied in~\cite{Bekaert:2017bpy} to the Type-A higher spin gravity (mention also earlier developments~\cite{Grigoriev:2006tt,Grigoriev:2012xg,Bekaert:2012vt,Bekaert:2013zya,Alkalaev:2014nsa}), in this work we made it more precise and used to construct a new higher spin theory rather than giving different forms of already existing ones.

There is a number of models closely related to the Type-A (Type-B) higher spin gravity that is dual to $\square \phi=0$ ($\slashed{\pl} \psi=0$) free CFT. When both CFT's are taken on an arbitrary background for single-trace operators, partition functions are
\begin{align*}
    e^{W_A[H_A]}&=\int D\phi\,D\bar\phi\, e^{ -\int \bar\phi H_A(x,\pl)\phi }\,, &
    e^{W_B[H_B]}&=\int D\psi\,D\bar\psi\, e^{ -\int \bar\psi H_B(x,\pl,\gamma) \psi}\,.
\end{align*}
This formal expressions require certain zeroth order value (vacuum) for $H_{A,B}$ to have well-defined expansions. The simplest choice is $H_A=-\pl^2+...$ and $H_B=i\slashed{\pl}+...$. In these two cases the symmetries of these backgrounds correspond to higher spin currents constructed out of free boson, $\square \phi=0$, and free fermion, $\slashed{\pl} \psi=0$, respectively. An advantage of such general treatment is that there can be other vacua. Indeed, one can expand over $H_A=(-\pl^2)^k$ and  $H_B=i(-\pl^2)^k\slashed{\pl}$ \cite{Bekaert:2013zya,Grigoriev:2014kpa}. The symmetries of these vacua correspond to non-unitary free CFT's 
$\square^k \phi=0$, and $\square^k\slashed{\pl} \psi=0$, see e.g.
\cite{Skvortsov:2006at,Alkalaev:2014nsa,Joung:2015jza,Gunaydin:2016amv, Brust:2016xif,Basile:2014wua}. Therefore, the duals of these non-unitary CFT's correspond to different vacua for $F_i$ fields.

An important property of our method is that the off-shell theory of background fields on the boundary and the on-shell theory in the bulk merely arise as two different vacua in one and the same system, giving an explicit realization of the duality. It is important to stress that our method is general and can be applied to any CFT (not necessarily a free one) and it would be interesting to apply it to other dualities. If the symmetries of background fields do not fix the effective action unambiguously we should find the corresponding ambiguity in the reconstruction procedure.

It is remarkable that there exists such a simple relation between the theory of background fields and the dual higher spin theory. It may not be obvious at the moment why the direct computation of correlation functions should give the correct result $W[\tildef]$ that would prove the duality. Nevertheless, general, but indirect, proof is possible: it can be shown that the boundary values of the $AdS_{d+1}$ fields are exactly the source $\tildef$ together with the full gauge transformations that are known to completely fix $W[\tildef]$, \cite{Maldacena:2011jn,Alba:2013yda,Boulanger:2013zza,Alba:2015upa}.\footnote{Let us note that the uniqueness result for $W[\tildef]$ has been obtained under certain assumptions. Some of these assumptions correspond, in our language, to the choice of a particular vacuum: the one given by a unitary CFT. We expect that unitarity can be relaxed and all CFT's in $d>2$ with higher spin currents are free. We stress that choice of a vacuum is important and $\mathrm{tr} \log H$ is just some formal expression that should be understood as being expanded over some vacuum. Also, the uniqueness result assumes one free parameter, the number of fields $N$. Large-$N$ expansion is trivial in free CFT's: $N$ just gives the relative ratio between connected and disconnected contributions to the full correlation functions. On the $AdS_{d+1}$ side it is important to assume the bulk coupling constant, which is of order $1/N$, to be small, the latter justifies quasi-classical expansion. } Still, it is worth mentioning, as we discuss below, that any direct computation of correlation functions is likely to face some difficulties due to the lack of understanding of micro-locality in the context of higher spin theories. Therefore, our main assumption here is that there exists some scheme in the bulk that does not destroy a near boundary analysis of symmetries.

One can also try to understand the relation between the theory of background fields and dual higher spin theory from a more formal perspective. Indeed, these two theories are closely related because the theory on the boundary is precisely the theory of boundary values for the bulk one. At the same time a reparametrization invariant gauge theory is entirely determined by the on-shell gauge transformation (more precisely, the BRST differential defined on the stationary surface extended by ghost variables. See~\cite{Barnich:2010sw} for more detail.) but this is precisely the data encoded in the boundary theory. This gives a somewhat heuristic argument on how the bulk theory can be systematically reconstructed from the known theory of boundary values. Let us also mention that this approach is somewhat complementary to the holographic reconstruction where the vertices in the bulk are built in such a way that the correlation functions of a given CFT are reproduced through the usual AdS/CFT prescription \cite{Bekaert:2014cea,Bekaert:2015tva,Sleight:2016dba}, which is still perturbative.

A word of warning is needed about feasibility of usual field theory computations in higher spin gravities, including the one presented here. A careful analysis of the quartic interactions in the simplest Type-A higher spin gravity has revealed \cite{Bekaert:2015tva,Sleight:2017pcz,Ponomarev:2017qab} that starting from the quartic order the interactions of higher spin fields become too non-local for usual field methods to work without thinking and further specification of how to deal with non-localities is needed. For example, sum over derivatives may not commute with sum over spins, etc. That the sum over spins is not convergent and requires regularization was also observed for one-loop determinants, see e.g. \cite{Giombi:2013fka,Giombi:2014yra,Bae:2016rgm,Giombi:2016pvg,Gunaydin:2016amv,Skvortsov:2017ldz}. Also, higher spin gravities should emerge in tensionless limits of string theory, which is unlikely to give an ordinary field theory. 

Therefore, we are led to think that higher spin gravities are more stringy than it has been previously thought. While our approach shows that higher spin symmetries can fully be taken into account and a background independent description of the theory can be constructed, an attempt to dissect the interactions into sums over spins and derivatives will surely fail to give any meaningful result (except for the few terms at the lowest order) unless a proper regularization is found. An appropriate stringy way to deal with higher spin gravities is yet to be found. Nevertheless, as our method takes into account the full non-abelian gauge symmetry of boundary sources that fixes completely the effective action, we can argue that the AdS/CFT duality follows automatically at least at the tree level.  

An interesting feature of the approach advocated in the paper is that the models can be truncated to lower spins. This is certainly possible in the effective action as there is no prior need to introduce sources for anything that goes beyond the stress-tensor multiplet. However, once at least one higher spin source is introduced we will have to add all of them as to make the effective action gauge invariant. The initial system of equations is off-shell in this sense and truncation to lower spin sector is possible, which is not anymore possible once we go on-shell by factoring out configurations proportional to the fields themselves.

Finally, we would like to remind that in even dimensions the effective action $W[\tildef]$ has a $\log$-divergent part $W_{\log}[\tildef]$ to be identified with higher spin conformal anomaly \cite{Segal:2002gd,Tseytlin:2002gz}, see also \cite{Bekaert:2010ky,Grigoriev:2016bzl,Bonezzi:2017mwr}. By itself, this local part $W_{\log}[\tildef]$ can be used to define Type-B conformal higher spin theory on the boundary. Also, the same $W_{\log}[\tildef]$ should be obtained from the effective action on the $AdS_{d+1}$ side.

%%%%%%%%%%%%%%%%%%%%%%%%%%%%%%%%%%%%%%%%%%%%%%%%%%%%%%%%%%%%%
\section*{Acknowledgments}
\label{sec:Aknowledgements}
%%%%%%%%%%%%%%%%%%%%%%%%%%%%%%%%%%%%%%%%%%%%%%%%%%%%%%%%%%%%%
We are grateful to Kostya Alkalaev for collaboration at the early stage of this project. We are also grateful to Xavier Bekaert, Ivo Sachs, Alexey Sharapov, Cedric Troessaert, Arkady Tseytlin, Sasha Zhiboedov for useful discussions. 
The work of M.G. was supported by the DFG Transregional Collaborative Research Centre TRR 33 and the DFG cluster of excellence ``Origin and Structure of the Universe'' and by the RFBR grant 18-02-01024. The work of E.S. was supported by the Russian Science Foundation grant 14-42-00047 in association with the Lebedev Physical Institute. 

%%%%%%%%%%%%%%%%%%%%%%%%%%%%%%%%%%%%%%%%%%%%%%%%%%%%%%%%%%%%%
\begin{appendix}
\renewcommand{\thesection}{\Alph{section}}
\renewcommand{\theequation}{\Alph{section}.\arabic{equation}}
\setcounter{equation}{0}\setcounter{section}{0}
%%%%%%%%%%%%%%%%%%%%%%%%%%%%%%%%%%%%%%%%%%%%%%%%%%%%%%%%%%%%%

%%%%%%%%%%%%%%%%%%%%%%%%%%%%%%%%%%%%%%%%%%%%%%%%%%%%%%%%%%%%%
\section{Going to Symmetric Base}
\setcounter{equation}{0}
\label{sec:symmbase}
%%%%%%%%%%%%%%%%%%%%%%%%%%%%%%%%%%%%%%%%%%%%%%%%%%%%%%%%%%%%%

Here we prove the ambient space version of the statement. The version we need in section~\bref{sec:fermion-mink} is obtained by relaxing the constraints \eqref{const-1}-\eqref{const-13-gs} and assuming all fields and parameters totally traceless.

Consider the ambient space system~\eqref{linearized-hook}, \eqref{linearized-hook-eoms+traces}. One can check that the gauge condition  $(\mph \cdot \pl_P) f_+=0$ is reachable. The gauge condition simply says that $f_+$ is associated to Young diagram of hook shape in the antisymmetric basis: one can choose the component fields to be anti-symmetric in the indices corresponding to the first column and symmetric in the rest of the indices. Then, the Young symmetry condition is exactly $(\mph \cdot \pl_P) f_+=0$

Indeed, if $(\mph\cdot \d_P) f_+=\alpha$ let us take $\xi=-\frac{1}{N}  \alpha$, where $N=\mph\cdot \d_\mph+P\cdot \d_P$. Note that $\xi$ satisfy the constraints above.
 It follows,
\begin{equation}
\mph\cdot \d_P (s- \frac{1}{N}(p\cdot \d_\mph-\mph\cdot \d_X)\alpha)=
\alpha-\frac{1}{N}N\alpha+\frac{1}{N}
(\mph\cdot \d_P)(\mph\cdot \d_X) \alpha\,,
\end{equation}
where we used $(\mph\cdot \d_P) \alpha=0$. Iterating the procedure we arrive at the gauge condition $(\mph\cdot \d_P) f_+=0$.

In this gauge one can employ the Poincare Lemma in the $\mph,P$-space and express $f_+$ as $f_+=(\mph\cdot\d_P) \psi$ in terms of the new generating function $\psi$. Note that $\psi$ can be assumed to satisfy $(P\cdot \d_\mph) \psi=0$, i.e. the Young condition in symmetric basis.

Let us analyze the residual symmetries in some more detail: let us decompose the space of $\cH$ of polynomials in $\mph,p$ (without $1$) as a direct sum $\cH=\cH^\parallel\oplus \cH^{\perp}$ where $\cH^\parallel=\ker(\mph\cdot\d_P)$ and $\cH^\perp=\ker(P\cdot\d_\mph)$.
Decompose the gauge variation as:
\begin{equation}
\label{gauge-var}
    \mph\cdot\d_X \xi^\parallel+(\mph\cdot\d_X \xi^\perp)^\parallel+
    (\mph\cdot\d_X \xi^\perp)^\perp-P\cdot\d_\mph \xi^\parallel\,. 
\end{equation}
Vanishing of the $\perp$ contribution implies $
(\mph\cdot\d_X \xi^\perp)^\perp-P\cdot\d_\mph \xi^\parallel =0
$. 
It is clear that this determines $\xi^\parallel$ uniquely. Moreover, for $\xi^\perp$ homogeneous in $p,\mph$, resulting $\xi^\parallel$ is proportional to
$(\mph\cdot\d_P)( \mph\cdot\d_X) \xi^\perp$ and hence the first term in the gauge variation~\eqref{gauge-var} vanish thanks to nilpotency of $\mph\cdot \d_X$. Finally, the gauge variation takes the form:
\begin{equation}
    \delta f_+=(\mph\cdot \d_X \xi^\perp)^\parallel
\end{equation}
The variation of $\psi$ is then proportional to:
\begin{equation}
P\cdot \d_\mph((\mph\cdot \d_X \xi^\perp)^\parallel)=
P\cdot \d_\mph((\mph\cdot \d_X \xi^\perp)=
P\cdot\d_X \xi^\perp\,,
\end{equation}
giving the standard gauge law in symmetric basis. Finally, analyzing constraints on $\psi,\xi^\perp$ one finds that these are precisely~\eqref{amb-hook}, where $f_S=\psi$ and $\xi_S=\xi^\perp$. Moreover, if in addition $f_+,\xi$ are subject to the analog of~\eqref{linearized-hook-eoms+traces} then 
the resulting  $f_S,\xi_S$ also satisfy~\eqref{eoms+traces}.

%%%%%%%%%%%%%%%%%%%%%%%%%%%%%%%%%%%%%%%%%%%%%%%%%%%%%%%%%%%%%
\end{appendix}
%%%%%%%%%%%%%%%%%%%%%%%%%%%%%%%%%%%%%%%%%%%%%%%%%%%%%%%%%%%%%
\setstretch{1.0}
\providecommand{\href}[2]{#2}\begingroup\raggedright\endgroup

\end{document}